\documentclass[aps,prd,twocolumn,floatfix,nofootinbib,showpacs,superscriptaddress,tightenlines]{revtex4}

\usepackage{amsmath}
\usepackage{lipsum}
\usepackage{amssymb}
\usepackage{amsthm}
\usepackage{bbold}
\usepackage{dcolumn}
\usepackage{epsfig}
\usepackage{graphics}
\usepackage{graphicx}
\usepackage{longtable}
\usepackage{color}
\usepackage{epstopdf}
\usepackage{xspace}
\usepackage{cancel}
\usepackage{nicefrac}
\usepackage[colorlinks=true, pdfstartview=FitV, linkcolor=purple, citecolor= purple,urlcolor=blue]{hyperref}

\definecolor{darkgreen}{rgb}{0,0.5,0}
\definecolor{purple}{rgb}{0.5,0,0.5}
\definecolor{nblue}{rgb}{0.0,0.0,0.50}
\definecolor{scarlet}{rgb}{1.0,0.2,0}

\begin{document}

\title{Transverse Takahashi Identities and Their Implications for Gauge Independent Dynamical Chiral Symmetry Breaking}

\author{L. Albino}
\affiliation{Instituto de F\'{i}sica y Matem\'aticas, Universidad
Michoacana de San Nicol\'as de Hidalgo, Morelia, Michoac\'an
58040, M\'{e}xico.}
\author{A. Bashir}
\affiliation{Instituto de F\'{i}sica y Matem\'aticas, Universidad
Michoacana de San Nicol\'as de Hidalgo, Morelia, Michoac\'an
58040, M\'{e}xico.}
\author{L.X. Guti\'errez Guerrero}
\affiliation{CONACyT-Mesoamerican Centre for Theoretical Physics, Universidad Aut\'onoma de Chiapas, Carretera Zapata Km. 4, Real del Bosque (Ter\'an), Tuxtla Guti\'errez 29040, Chiapas, M\'{e}xico.}
\author{B. El Bennich}
\affiliation{Laborat\'orio de F\'isica Te\'orica e Computacional,
Universidade Cruzeiro do Sul, Rua Galv$\tilde{\rm a}$o Bueno 868,
01506- 000 S$\tilde{\rm a}$o Paulo, SP, Brazil.}
\author{E. Rojas}
\affiliation{Departamento de F\'isica, Universidad de Nari$\tilde{n}$o,
A.A. 1175, San Juan de Pasto, Colombia.}

\date{\today}

\begin{abstract}
%
%Non-perturbative studies of the Schwinger-Dyson equations require
%their infinite, coupled tower to be truncated in order to reduce
%them to a practically solvable set. A usual procedure is to make
%an \textit{Ansatz} for the three-point fermion-photon vertex. Quantum
%electrodynamics is an illuminating example in which we can
%study the constraints of gauge invariance, namely, the
%Ward-Fradkin-Green-Takahashi identities, which determine
%the so called longitudinal part of this vertex and the
%Landau-Khalatnikov-Fradkin transformations, which are concerned with its
%component transverse to the photon momentum. In hadronic observables, such as form factors, where the external probe is electromagnetic, this vertex plays an
%important role in theoretical predictions.
In this article, we employ transverse Takahashi identities to impose valuable non-perturbative constraints on
the transverse part of the fermion-photon vertex in terms of new form factors, the so called $Y_i$ functions. We show
that the implementation of these identities is crucial in ensuring
the correct local gauge transformation of the fermion propagator
and its multiplicative renormalizability. Our construction incorporates
the correct symmetry properties of the $Y_i$ under charge conjugation operation
as well as their
well-known one-loop expansion in the asymptotic configuration of incoming
and outgoing momenta. Furthermore, we make an explicit
analysis of various existing constructions of this vertex
against the demands of transverse Takahashi identities and the previously established key features of quantum electrodynamics, such as gauge invariance of the critical coupling above which chiral symmetry is dynamically broken.
We construct a simple example in its quenched version and compute the mass function
as we vary the coupling strength and also calculate the corresponding anomalous dimensions $\gamma_m$.
There is an excellent fit to the Miransky scalling law and we find $\gamma_m=1$ rather
naturally in accordance with some earlier results in literature, using arguments based on
Cornwall-Jackiw-Tomboulis effective potential technique. Moreover, we numerically confirm the gauge invariance of this critical coupling.

\end{abstract}


\maketitle

\date{\today}

%%%%%%%%%%%%%%%%%%%%%%%%%%%%%%%%%%%%%%%%%%%%%%%%%%%%%%%%%%%%%%%
%%%%%%%%%%%%%%%%%%%%%%%%%%%%%%%%%%%%%%%%%%%%%%%%%%%%%%%%%%%%%%%
%%%%%%%%%%%%%%%%%%%%      INTRODUCTION     %%%%%%%%%%%%%%%%%%%%
%%%%%%%%%%%%%%%%%%%%%%%%%%%%%%%%%%%%%%%%%%%%%%%%%%%%%%%%%%%%%%%
%%%%%%%%%%%%%%%%%%%%%%%%%%%%%%%%%%%%%%%%%%%%%%%%%%%%%%%%%%%%%%%

\section{Introduction}
\label{SECTION Introduction}

A quantum field theory (QFT) can be considered completely solved
if we are able to compute its full set of $n$-point Green
functions. However, these Green functions are infinite in number.
Moreover, these are all intertwined through highly non-linear
coupled integral equations, known as the Schwinger-Dyson equations
(SDEs). A brute force method to compute them is a wild goose
chase.

We believe that a satisfactory determination of relevant
physical observables through a systematic truncation scheme for
this infinite tower of equations is achievable if we preserve the
key features and symmetries of the underlying theory. Perturbation theory (PT)
provides an excellent example of such an approximation scheme.
However, when the interaction strength grows and can no longer be
used as a perturbative expansion parameter, one resorts to
truncations which need to be carefully constructed in order to
retain the essential features of the original theory, while maintaining
contact with experimental data at the same time. Quantum
chromodynamics (QCD) is a realization of this scenario in its
infrared domain. Considerable progress has been made in the last
decades to study its first few Green functions, e.g.,
the gluon propagator~\cite{Boucaud:2008ky,Aguilar:2008xm,Pennington:2011xs,Ayala:2012pb,Aguilar:2012rz,Strauss:2012dg,Bashir:2013zha,Huber:2015ria}
and the quark-gluon vertex~\cite{Alkofer:2008tt,Chang:2009zb,Kizilersu:2009kg,Bashir:2011dp,Hopfer:2013np,Rojas:2013tza,Aguilar:2014lha,Pelaez:2015tba,Williams:2015cvx,Aguilar:2018epe}
whose knowledge consequently provides predictions for QCD and hadron physics, e.g.,~\cite{Alkofer:2008et,Chang:2013nia,Williams:2014iea,Raya:2015gva,Gomez-Rocha:2016cji}; also see
reviews~\cite{Bashir:2004mu,Bashir:2012fs,Aznauryan:2012ba,Eichmann:2016yit,Huber:2018ned} and references therein.

In several hadronic physics studies, such as electromagnetic and transition form factors~\cite{Chang:2013nia,Raya:2015gva,Raya:2016yuj,Weil:2017knt,Eichmann:2017wil,Ding:2018xwy}, probes are generally electromagnetic in nature and many SDE calculations crucially rely on how photons interact with quarks.
Thus, quantum electrodynamics (QED) serves as a useful platform to
study SDE truncations and provide improvements to
preserve its key features, such as its gauge
invariance, renormalizability and the recuperation of the well-known
S-matrix perturbative expansion for its Green functions in the weak
coupling regime, which it maintains at all accessible energies.
In particular, study of the fermion propagator
in QED generally amounts to requiring a physically
meaningful and reliable \textit{Ansatz} for the three-point
fermion-photon vertex. Gauge invariance provides an essential
ingredient in this connection. The gauge technique of Salam, Delbourgo and collaborators
was developed to solve the constraints of the well-known Ward-Fradkin-Green-Takahashi identity
(WFGTI)~\cite{Ward:1950xp,Fradkin:1955jr,Green:1953te,Takahashi:1957xn}, writing the Green functions
in terms of spectral representations,~\cite{Salam:1964zk,Delbourgo:1977jc,Delbourgo:1978bu}.
However, such approach, despite its elegant and formal results,~\cite{Delbourgo:1980vc,Delbourgo:1984gu}, is not amicable to straightforward
computations,~\cite{Rembiesa:1986mz}. The WFGTI
allows us to expand out the vertex in terms of a well-constrained
longitudinal part~\cite{Ball:1980ay} and an undetermined
transverse part. Several efforts alternative to the gauge technique start
from making an \textit{Ansatz} for this latter part and proceed from thereon.

A natural question to ask is {\em if the transverse part can be constrained
through any other symmetry principle?}

Whereas the usual WFGTI relates the divergence of the three-point
fermion-photon vertex to the inverse fermion propagator, there exist
transverse Takahashi identities (TTI) which play a similar role
for the curl of the fermion-photon
vertex~\cite{Takahashi:1985yz,Kondo:1996xn,He:2000we,He:2006my,He:2007zza}.
However, in addition to the inverse fermion propagator and the
vector vertex, these identities also bring into play a non-local axial-vector
vertex as well as new inhomogeneous tensor and axial-tensor
vertices. Consequently, TTI are richer and more complicated in
their structure. In past, they have been verified to
one-loop order,~\cite{Pennington:2005mw,He:2006ce}. More recently, practical
implications of TTI have been
investigated in~\cite{Qin:2013mta,Qin:2014vya} to get insight into
the non-perturbative forms of vector and axial-vector vertices.

In this article, we intend to study constraints of TTI on
the transverse part of the fermion-photon vertex. Note that
TTI do not modify the usual WFGTI in any way. However, we realize
that they are crucially connected to another consequence of local
gauge covariance, namely, Landau-Khalatnikov-Fradkin
transformations (LKFT), derived
in~\cite{Landau:1955zz,Fradkin:1955jr,Johnson:1959zz,Zumino:1959wt}.
LKFT are a well defined set of transformations which describe
the response of the Green functions to an arbitrary gauge
transformation. These transformations leave the SDEs and the WFGTI
form-invariant. LKFT potentially play an important role in
imposing valuable constraints on the fermion-photon vertex and
obtaining gauge invariant chiral symmetry breaking, see for
example
Refs.~\cite{Burden:1993gy,Bashir:1999bd,Bashir:2000ur,Bashir:2002sp,Bashir:2004hh,Fischer:2004nq,Bashir:2005wt,Bashir:2006ga,Bashir:2007qq,Bashir:2008ej,Bashir:2009fv,Ahmadiniaz:2015kfq,Ahmad:2016dsb,Pennington:2016vxv,Jia:2016wyu,Jia:2016udu}.
More recently, these transformations have also been derived for
QCD~\cite{DeMeerleer:2018txc,Aslam:2015nia}.

Both the TTI and the LKFT (through the multiplicative
renormalizability (MR) of the fermion propagator) constrain the
transverse fermion-photon vertex. Therefore, it is reasonable to
seek a combined constraint which would help us converge on pinning
down this elusive part of the vertex. The fact that MR constrains
the transverse vertex has already been known for some time
~\cite{Parker:1984gx,Delbourgo:1984gu,Cornwall:1981zr,King:1982mk}.
Later works in the literature involving similar
considerations in constructing a refined fermion-photon vertex
can be found in~\cite{Curtis:1990zs,Curtis:1991fb,Curtis:1993py,Dong:1994jr,Bashir:1994az,Bashir:1995qr,Bashir:1997qt,Kizilersu:2009kg,Fernandez-Rangel:2016zac}.

An important issue relevant to our current work concerns the usage of the
TTI-constrained vertex to study dynamical chiral symmetry breaking (DCSB)
or dynamical fermions mass generation as a consequence of enhanced interaction
strength. This is a strictly non-perturbative phenomenon and a transcendental topic
in QCD, where it induces measurable effects in numerous hadron observables.
Therefore, physically meaningful truncations of QCD's SDEs demand
incorporation of DCSB through the relevant Green functions, in
particular the quark propagator and the quark-gluon vertex.
Regarding the latter, valuable progress has been made both in
lattice,~\cite{Skullerud:2003qu,Skullerud:2004gp,Kizilersu:2006et,Oliveira:2016muq,Sternbeck:2017ntv,Oliveira:2018fkj,Oliveira:2018ukh} and continuum studies. However, due to the non-abelian
nature of QCD, investigating the impact of DCSB on the quark-gluon
vertex, and vice versa, from the first principles, is still a
theoretical challenge. A thorough investigation of the fermion-photon
vertex and chiral symmetry breaking in QED is likely to provide a
bench mark for the corresponding studies in QCD.

Although QED manifests a perturbative behavior at all observable scales,
an intense background electromagnetic field can trigger a
transition from perturbative to non-perturbative dynamics, the well-known
{\em magnetic catalysis}, see for example~\cite{Gusynin:1994xp,Gusynin:1995gt,Lee:1997zj,Hong:1997uw,Ferrer:2000ed,Ayala:2006sv,Rojas:2008sg,Watson:2013ghq,Mueller:2015fka}. Even a toy QED with an artificially scaled up coupling
exhibits this phenomenon. Such a phase transition has long
been studied. It is characterized by a critical coupling,
$\alpha_c$, above which DCSB takes place,
see~\cite{Curtis:1990zs,Bashir:2011dp,Kizilersu:2014ela} and references therein. Since this
critical coupling corresponds to a recognizable phase transition, it is
considered to be a physical observable, and hence a gauge
invariant parameter. This independence of $\alpha_c$ on the gauge
parameter has long been used as a further requirement to constrain the
transverse vertex~\cite{Bashir:2011dp}, and we follow
this argument in the present article.

The TTI connect the transverse structure of the
fermion-photon vertex to a set of unknown scalar functions
$Y_i$ related to a non-local axial-tensor mentioned before.
MR of the electron propagator implies that these
functions cannot be ignored. Instead, MR constrains their
form. This procedure involves an unknown function $W(x)$ of a
dimensionless ratio $x$ of the incoming and outgoing fermion
momenta. It satisfies an integral constraint which guarantees
MR of the electron propagator in the leading logarithm
approximation (LLA). Implementing charge conjugation symmetry
on the integration kernels involved in the fermion propagator SDE,
it is possible to parameterize $Y_i$ in terms of one single
scalar, \textit{a priori} unknown function ${\cal{T}}(k^2,p^2)$,
which encodes the effect of the fully-dressed fermion-photon vertex on the fermion propagator. This general procedure
fixes three of the $Y$-functions. An additional constraint comes
from demanding gauge independent chiral symmetry breaking. For
the so called quenched approximation, it yields a self-consistent
solution for ${\cal{T}}(k^2,p^2)$.

In this article, we work in Euclidean space. Thus, for
$\gamma$-matrices we have: $\left\{ \gamma_{\mu},\gamma_{\nu}
\right\} = 2 \delta_{\mu\nu}$ and
$\gamma^{\dagger}_{\mu}=\gamma_{\mu}$, where $\delta_{\mu\nu}$ is
the Euclidean metric. Furthermore, we define
$\gamma_{5}=\gamma_{4} \gamma_{1} \gamma_{2} \gamma_{3}$, with
$\hbox{Tr}\left[ \gamma_{5} \gamma_{\mu} \gamma_{\nu}
\gamma_{\alpha} \gamma_{\beta} \right] = -4 \epsilon_{\mu \nu
\alpha \beta}$.

This paper is organized as follows: in Section~\ref{SECTION Vertex
decomposition}, we review the WFGTI for the three-point vertex in
QED, define the longitudinal vertex, write down the transverse part
in a general basis, following~\cite{Ball:1980ay}, and highlight the symmetry
properties of the transverse form factors under charge conjugation operation.
In Section~\ref{SECTION
Transverse TI}, we introduce abelian TTI for the vertex and
expand out the transverse form factors in terms of the
$Y$-functions. We then invert these relations and impose a perturbative
constraint on $Y_{i}$ in the asymptotic limit of $k^2 \gg p^2$.
In Section~\ref{SECTION Gap equation}, SDE
for the fermion propagator is presented. We write it in terms of the $Y_{i}$ functions and discuss the quenched approximation.
In Section~\ref{SECTION MR constraints},
We study the
requirement of MR and the power law solution for the wave function
renormalization of the fermion propagator within the LLA. We show how the requirement of MR
imposes an integral constraint on the form
factors of the transverse vertex, in terms of the function $W(x)$. In Section~\ref{SECTION
Example}, we discuss a few examples illustrating the need and
importance of the $Y$-functions. In Section~\ref{SECTION Gauge independent DCSB},
we construct simple examples to study DCSB and see how it naturally
incorporates gauge independence of the critical coupling $\alpha_c$.
In Section~\ref{SECTION Results and Conclusions}, we present our conclusions
and discuss prospects for future research.

%%%%%%%%%%%%%%%%%%%%%%%%%%%%%%%%%%%%%%%%%%%%%%%%%%%%%%%%%%%%%%%
%%%%%%%%%%%%%%%%%%%%%%%%%%%%%%%%%%%%%%%%%%%%%%%%%%%%%%%%%%%%%%%
%%%%%%%%%%%%%%%%%     VERTEX DECOMPOSITION     %%%%%%%%%%%%%%%%
%%%%%%%%%%%%%%%%%%%%%%%%%%%%%%%%%%%%%%%%%%%%%%%%%%%%%%%%%%%%%%%
%%%%%%%%%%%%%%%%%%%%%%%%%%%%%%%%%%%%%%%%%%%%%%%%%%%%%%%%%%%%%%%

\section{Vertex decomposition}
\label{SECTION Vertex decomposition}

\begin{figure}[!ht]
    \centering
    \includegraphics[scale=.5]{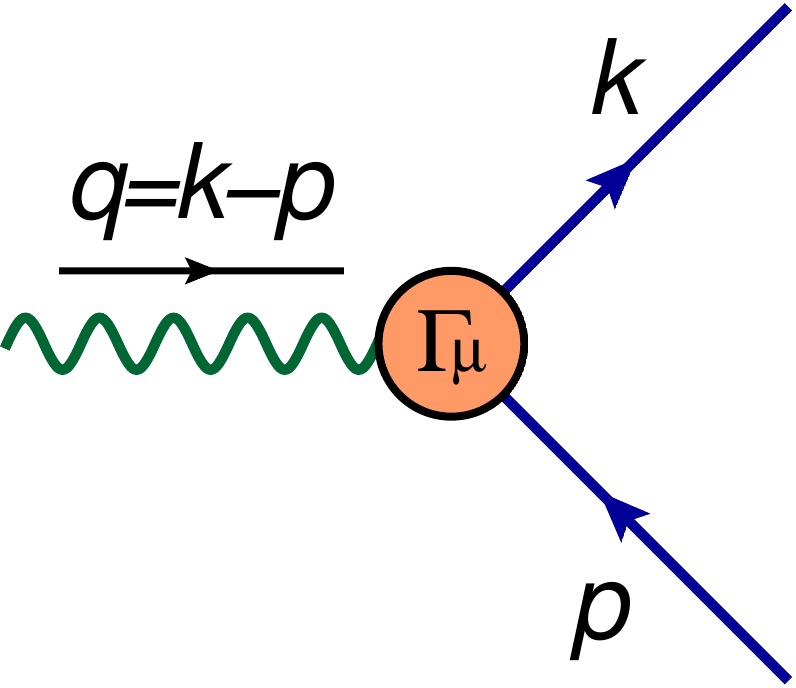}
    \caption{Diagrammatic representation of the full three-point vertex
$\Gamma_{\mu}(k,p)$, with momentum flow indicated.}
    \label{Vertice3PuntosLuis}
\end{figure}

In its general decomposition, the three-point fermion-boson vertex
can be
written in terms of 12 independent spin structures. For the kinematical
configuration of Fig.~\ref{Vertice3PuntosLuis}, the WFGTI
associated with this vertex takes the form
\begin{equation}
i q_{\mu}\Gamma_{\mu}(k,p)=S^{-1}(k)-S^{-1}(p) \,, \label{WGTI for the 3-point vertex}
\end{equation}
where $q=k-p$. This identity allows us to split the vertex as
a sum of \textit{longitudinal} and \textit{transverse} components,
as suggested by Ball and Chiu~\cite{Ball:1980ay}:
\begin{equation}
\Gamma_{\mu}(k,p)=\Gamma^{L}_{\mu}(k,p)+\Gamma^{T}_{\mu}(k,p) \,. \label{Ball-Chiu vertex decomposition}
\end{equation}
The longitudinal part $\Gamma^{L}_{\mu}(k,p)$ alone satisfies the WFGTI (\ref{WGTI for the 3-point vertex}), and consumes four of the twelve independent spin structures (one of them is zero in QED), so that,~\cite{Ball:1980ay}:
\begin{eqnarray}
\Gamma^{L}_{\mu}(k,p)= a(k^{2},p^{2}) \gamma_{\mu} +
\frac{b(k^{2},p^{2})}{2} t_{\mu} \gamma \cdot t
 - i  c(k^{2},p^{2}) t_{\mu} \label{Longitudinal vertex
decomposition}
\end{eqnarray}
with $t=k+p$, and
\begin{eqnarray}
a(k^{2},p^{2}) &=& \frac{1}{2} \left[
\frac{1}{F(k^{2},\Lambda^{2})} +
\frac{1}{F(p^{2},\Lambda^{2})} \right] \,, \nonumber \\
b(k^{2},p^{2}) &=& \left[ \frac{1}{F(k^{2},\Lambda^{2})} -
\frac{1}{F(p^{2},\Lambda^{2})} \right] \frac{1}{k^{2}-p^{2}} \,, \nonumber \\
c(k^{2},p^{2}) &=& \left[ \frac{ {\cal{M}}(k^{2},\Lambda^{2})
}{F(k^{2},\Lambda^{2})} - \frac{ {\cal{M}}(p^{2},\Lambda^{2})
}{F(p^{2},\Lambda^{2})} \right] \frac{1}{k^{2}-p^{2}} \,, \label{longitudinal coefficients definitions}
\end{eqnarray}
where $\Lambda$ is an ultraviolet cut-off regulator. ${\cal{M}}$
and $F$ are the mass function and the wave function
renormalization, respectively, related to the fermion propagator $S(k)$
through
\begin{equation}
S(k) = \frac{F(k^{2},\Lambda^{2})}{i \gamma \cdot k +
{\cal{M}}(k^{2},\Lambda^{2})} \,. \label{fermion propagator
definition}
\end{equation}
At the tree level, $F(k^{2},\Lambda^{2})=1$ and
${\cal{M}}(k^{2},\Lambda^{2})=m_0$, where $m_0$ is the bare mass
of the fermion.

The transverse part $\Gamma^{T}_{\mu}(k,p)$ of the vertex
decomposition (\ref{Ball-Chiu vertex decomposition}), which
remains undetermined by the WFGTI, is naturally constrained by
\begin{equation}
 q_{\mu}\Gamma^{T}_{\mu}(k,p)=0 \,. \label{transverse part
definition}
\end{equation}
In general, the ultraviolet finite transverse vertex can be
expanded out in terms of 8 vector structures, and their
corresponding scalar form factors $\tau_{i}(k,p)$~\cite{Ball:1980ay}:
\begin{equation}
\Gamma^{T}_{\mu}(k,p) = \sum_{i=1}^{8} \tau_{i}(k,p)
T^{i}_{\mu}(k,p)\,. \label{transverse vertex structure}
\end{equation}
%where the simplifying notation $(k,p)$ indicates a dependence on $k^2$, $p^2$ and $q^2$.
Moreover, for the kinematical configuration of Fig.~\ref{Vertice3PuntosLuis}, we define
\begin{eqnarray}
T^{1}_{\mu}(k,p) &=& i \left[ p_{\mu} (k \cdot q) -k_{\mu} (p
\cdot q) \right] \,, \nonumber \\
T^{2}_{\mu}(k,p) &=& \left[ p_{\mu} (k \cdot q) -k_{\mu} (p \cdot
q) \right] \gamma \cdot t \,, \nonumber \\
T^{3}_{\mu}(k,p) &=& q^{2} \gamma_{\mu} - q_{\mu} \gamma \cdot q \,, \nonumber \\
T^{4}_{\mu}(k,p) &=& i q^{2} \left[ \gamma_{\mu} \gamma \cdot t -
t_{\mu} \right] + 2 q_{\mu} p_{\nu} k_{\rho} \sigma_{\nu \rho}  \,, \nonumber \\
T^{5}_{\mu}(k,p) &=& \sigma_{\mu \nu} q_{\nu}  \,, \nonumber \\
T^{6}_{\mu}(k,p) &=& - \gamma_{\mu} \left( k^{2}-p^{2} \right) +
t_{\mu} \gamma \cdot q \,, \nonumber \\
T^{7}_{\mu}(k,p) &=&  \frac{i}{2} (k^{2}-p^{2}) \left[
\gamma_{\mu} \gamma \cdot t - t_{\mu} \right] + t_{\mu} p_{\nu}
k_{\rho} \sigma_{\nu  \rho}   \,, \nonumber \\
T^{8}_{\mu}(k,p) &=& -i \gamma_{\mu} p_{\nu} k_{\rho} \sigma_{\nu
 \rho} - p_{\mu} \gamma \cdot k + k_{\mu} \gamma \cdot p \,,
\label{transverse basis definition}
\end{eqnarray}
with
\begin{equation}
\sigma_{\nu \rho} = \frac{i}{2} \left[
\gamma_{\nu},\gamma_{\rho} \right] \,. \label{sigma definition}
\end{equation}
This basis is not exactly the one adopted in~\cite{Ball:1980ay}. We
choose to work with a modification of this initial basis which was put forward
in~\cite{Kizilersu:1995iz} and later employed in~\cite{Davydychev:2000rt} as well.
This latter choice ensures all transverse form factors of the vertex are independent of
any kinematic singularities in one-loop perturbation theory in an arbitrary covariant gauge.

As stated earlier in Section \ref{SECTION Introduction}, any
\textit{Ansatz} for the full vertex must have the same
transformation properties as the bare vertex under charge
conjugation operation. This requires all the $\tau_i$s in (\ref{transverse
vertex structure}) to be symmetric under the interchange $k
\leftrightarrow p$, except $\tau_{4}$ and $\tau_{6}$, which are
odd:
\begin{eqnarray}
&& \hspace{-.5cm} \tau_{i}(k,p)= \tau_{i}(p,k) \,, \hspace{.625cm}
i=1,2,3,5,7,8, \label{tau symmetric properties} \\
&& \hspace{-.5cm} \tau_{i}(k,p) = -\tau_{i}(p,k) \,, \quad i=4,6.
\label{tau antisymmetric properties}
\end{eqnarray}
From Eq.~(\ref{longitudinal coefficients definitions}), it is
obvious that $a(k^{2},p^{2})$, $b(k^{2},p^{2})$ and
$c(k^{2},p^{2})$ are symmetric under $k \leftrightarrow p$, as
they should be, in order to preserve the correct transformation
properties under charge conjugation operation for the full vertex.

Although the longitudinal scalar functions (\ref{longitudinal
coefficients definitions}) are fixed by the WFGTI (\ref{WGTI for
the 3-point vertex}), the transverse scalar functions in
decomposition (\ref{transverse vertex structure}) remain unknown.
In the next section, we introduce the TTIs for the three-point vertex
in QED, which provide a powerful tool in constructing
these non-perturbative transverse functions.

%%%%%%%%%%%%%%%%%%%%%%%%%%%%%%%%%%%%%%%%%%%%%%%%%%%%%%%%%%%%%%%
%%%%%%%%%%%%%%%%%%%%%%%%%%%%%%%%%%%%%%%%%%%%%%%%%%%%%%%%%%%%%%%
%%%%%%%%%%%%%%%%%%%     TAKAHASHI IDENT     %%%%%%%%%%%%%%%%%%%
%%%%%%%%%%%%%%%%%%%%%%%%%%%%%%%%%%%%%%%%%%%%%%%%%%%%%%%%%%%%%%%
%%%%%%%%%%%%%%%%%%%%%%%%%%%%%%%%%%%%%%%%%%%%%%%%%%%%%%%%%%%%%%%

\section{Transverse Takahashi identities}
\label{SECTION Transverse TI}

The TTIs for vector ($\Gamma_{\mu}$) and axial-vector
($\Gamma_{\mu}^{A}$) vertices in QED, related to a fermion with
bare mass $m_0$, read~\cite{Qin:2013mta}:
\begin{eqnarray}
q_{\mu} \Gamma_{\nu}(k,p) -q_{\nu} \Gamma_{\mu}(k,p)  &=&
S^{-1}(p) \sigma_{\mu\nu} + \sigma_{\mu\nu} S^{-1}(k) \nonumber \\
&& \hspace{-1.5 cm} +2 i m_0 \Gamma_{\mu\nu}(k,p) + t_{\alpha}
\epsilon_{\alpha \mu
\nu \beta} \Gamma^{A}_{\beta}(k,p) \nonumber \\
&& \hspace{-1.5 cm} + A^{V}_{\mu\nu}(k,p) \,, \label{T-WGTI Vector} \\
q_{\mu} \Gamma^{A}_{\nu}(k,p) -q_{\nu} \Gamma^{A}_{\mu}(k,p) &=&
S^{-1}(p) \sigma^{5}_{\mu\nu} - \sigma^{5}_{\mu\nu} S^{-1}(k) \nonumber \\
&& \hspace{-1.5 cm} + t_{\alpha} \epsilon_{\alpha \mu \nu \beta}
\Gamma_{\beta}(k,p) + V^{A}_{\mu\nu}(k,p) \,, \label{T-WGTI Axial}
\end{eqnarray}
where $\sigma^{5}_{\mu\nu}=\gamma_{5} \sigma_{\mu\nu}$, and
$\Gamma_{\mu\nu}(k,p)$ is an inhomogeneous tensor vertex. The last
two tensor structures in Eqs.~(\ref{T-WGTI Vector},\ref{T-WGTI
Axial}), $A^{V}_{\mu\nu}$ and $V^{A}_{\mu\nu}$, are related to
the momentum space expressions for non-local axial-vector and vector
vertices, whose definitions involve a
gauge-field-dependent line integral. These non-perturbative
identities are valid for any covariant gauge, and they do not have
explicit dependence on the covariant gauge parameter.

The vector and axial-vector TTIs are intricately coupled to each other via the
non-local terms $A^{V}_{\mu\nu}(k,p)$ and $V^{A}_{\mu\nu}(k,p)$,
which are complicated even at one-loop order,~\cite{Pennington:2005mw,He:2006ce}. Following the
procedure described in
Ref.~\cite{Qin:2013mta}, useful progress has been made to disentangle this
interdependence. In order
to project out transverse form factors from the TTIs,
Eqs.~(\ref{T-WGTI Vector},\ref{T-WGTI Axial}), it is convenient to
introduce the following tensors
\begin{eqnarray}
T^{1}_{\mu\nu} &=& \frac{1}{2} \epsilon_{\alpha \mu \nu \beta}  \, t_{\alpha} \, q_{\beta} \,,
\label{T1 mu nu} \\ \nonumber \\
T^{2}_{\mu\nu} &=& \frac{1}{2} \epsilon_{\alpha \mu \nu \beta}
 \, \gamma_{\alpha} \, q_{\beta} \,. \label{T2 mu nu}
\end{eqnarray}
By contracting the axial-vector identity (\ref{T-WGTI Axial}) with
tensors (\ref{T1 mu nu}) and (\ref{T2 mu nu}), the left-hand sides
of the resulting equations reduce to zero, while the right-hand
sides yield the following result:
\begin{eqnarray}
 q \cdot t \, t \cdot \Gamma(k,p) &=& T^{1}_{\mu\nu} \left[
 S^{-1}(p) \sigma^{5}_{\mu \nu} - \sigma^{5}_{\mu \nu} S^{-1}(k)
 \right] \nonumber \\
 && + t^{2} q \cdot \Gamma(k,p) + T^{1}_{\mu\nu} V_{\mu\nu}^{A}
 \,, \label{TTI 1} \\ \nonumber \\
 q \cdot t \gamma \cdot \Gamma(k,p) &=& T^{2}_{\mu\nu} \left[
 S^{-1}(p) \sigma^{5}_{\mu \nu} - \sigma^{5}_{\mu \nu} S^{-1}(k)
 \right] \nonumber \\
 && + \gamma \cdot t q \cdot \Gamma(k,p) + T^{2}_{\mu\nu} V_{\mu\nu}^{A}
 \,. \label{TTI 2}
\end{eqnarray}
These expressions only involve the vector vertex $\Gamma_{\mu}(k,p)$, and do
not contain explicit dependence on the fermion mass $m_0$.
Information about the axial-vector vertex $\Gamma^{A}_{\mu}(k,p)$
can be obtained through analogous procedure involving the
axial-vector TTI, Eq.~(\ref{T-WGTI Vector}). Although the terms
$T^{1}_{\mu\nu} V_{\mu\nu}^{A}$ and $T^{2}_{\mu\nu}
V_{\mu\nu}^{A}$ are still equally unknown, they are Lorentz scalar objects
and can thus be conveniently expressed as follows:
\begin{eqnarray}
iT^{1}_{\mu\nu} V_{\mu\nu}^{A} &=& \mathbf{I}_{D} Y_{1}(k,p) +
i(\gamma \cdot q) Y_{2}(k,p)
\nonumber \\
&& \hspace{-1cm} + i(\gamma \cdot t) Y_{3}(k,p) + \left[ \gamma
\cdot q, \gamma \cdot t \right] Y_{4}(k,p) \,, \label{TV1} \\ \nonumber \\
 iT^{2}_{\mu\nu} V_{\mu\nu}^{A} &=& i \mathbf{I}_{D} Y_{5}(k,p)
 + (\gamma \cdot q) Y_{6}(k,p) \nonumber \\
&& \hspace{-1cm} + (\gamma \cdot t) Y_{7}(k,p) +i \left[ \gamma
\cdot q, \gamma \cdot t \right] Y_{8}(k,p) \,, \label{TV2}
\end{eqnarray}
where $Y_{i}(k,p)$ are hitherto unconstrained scalar functions, and
$\mathbf{I}_{D}$ is the identity matrix. Projections of
Eqs.~(\ref{TTI 1},\ref{TTI 2}) lead to a set of eight linearly
independent, coupled linear equations that fix the eight
transverse scalar functions $\tau_i$ in terms of the $Y$-functions
defined via Eqs.~(\ref{TV1},\ref{TV2}).

From Eqs.~(\ref{transverse vertex structure},\ref{transverse basis
definition},\ref{TTI 1}-\ref{TV2}), it is possible to project out
the scalar form factors $\tau_i$:
\begin{eqnarray}
\tau_{1}(k,p) &=& - \frac{ Y_{1} }{ 2 (k^{2} - p^{2}) \nabla(k,p)
} \,,
\label{tau 1 from TTI} \\
\tau_{2}(k,p) &=& - \frac{Y_{5} - 3 Y_{3}}{ 4 (k^{2} - p^{2})
\nabla(k,p) } \,,
\label{tau 2 from TTI} \\
\tau_{3}(k,p) &=&  \frac{1}{2} b(k^2,p^2)
 + \frac{2 (k^{2}-p^{2}) Y_{2} - t^{2} (Y_{3} - Y_{5}) }{ 8
(k^{2} - p^{2}) \nabla(k,p) } \,, \qquad  \label{tau 3 from TTI} \\
\tau_{4}(k,p) &=&  - \frac{ (k^{2}-p^{2}) ( 6 Y_{4} + Y_{6}^{A} )
+ t^{2} Y_{7}^{S} }{ 8 (k^{2} - p^{2}) \nabla(k,p) } \,,
\label{tau 4 from TTI} \\
\tau_{5}(k,p) &=& - c(k^2,p^2) - \frac{2 Y_{4} + Y_{6}^{A}}{2 (k^{2}-p^{2})} \,, \label{tau 5
from TTI} \\
\tau_{6}(k,p) &=& \frac{2 q^{2} Y_{2} - (k^{2}-p^{2}) (Y_{3} -
Y_{5}) }{ 8 (k^{2} - p^{2}) \nabla(k,p) } \,,
\label{tau 6 from TTI} \\
\tau_{7}(k,p) &=& \frac{ q^{2} ( 6 Y_{4} + Y_{6}^{A} ) +
(k^{2}-p^{2}) Y_{7}^{S} }{ 4 (k^{2} - p^{2})
\nabla(k,p) } \,, \label{tau 7 from TTI} \\
\tau_{8}(k,p) &=& - b(k^2,p^2) - \frac{2 Y_{8}^{A}}{k^{2}-p^{2}} \,, \label{tau 8 from TTI}
\end{eqnarray}
where we have employed the obvious simplifying notation $Y_{i}\equiv Y_{i}(k,p)$.
Moreover, we have introduced the Gram determinant
\begin{equation}
\nabla(k,p) = k^{2} p^{2} - (k \cdot p)^{2} \,. \label{Gram
determinant}
\end{equation}
In addition, the vertex transformation properties under charge conjugation determine the symmetry properties
of the $Y$-functions:
\begin{eqnarray}
&& \hspace{-1cm} Y_{i}(k,p) = Y_{i}(p,k) \,, \hspace{.57cm}
i=2,6^{S},7^{S},8^{S},
\label{Y-functions symmetric properties} \\
&& \hspace{-1cm} Y_{i}(k,p) = -Y_{i}(p,k) \,, \quad
i=1,3,4,5,6^{A},7^{A},8^{A}, \label{Y-functions antisymmetric
properties}
\end{eqnarray}
where we conveniently introduce the decomposition
\begin{equation}
Y_{i}(k,p) = Y_{i}^{S}(k,p) + Y^{A}_{i}(k,p) \,, \label{Y8
symmetric-antisymmetric definition}
\end{equation}
for $i=6,7,8$, where the superscripts $S$ and $A$ stand for the
symmetric and antisymmetric parts of the corresponding $Y_{i}$s,
under $k \leftrightarrow p$. Note that in Eqs. (\ref{tau 1 from
TTI}-\ref{tau 8 from TTI}), there is no contribution of
$Y_{6}^{S}$, $Y_{7}^{A}$ and $Y_{8}^{S}$. This is a consequence of
the properties (\ref{tau symmetric properties}) and (\ref{tau
antisymmetric properties}), which entail
\begin{eqnarray}
Y_{6}^{S}(k,p) &=& - \frac{(k^{2}-p^{2}) Y_{1}(k,p)}{4
\nabla(k,p)} \,, \label{Y6 Sym in terms of taus} \\
Y_{7}^{A}(k,p) &=& \frac{q^{2} Y_{1}(k,p)}{4 \nabla(k,p)} \,,
\label{Y7 Antisym in terms of taus} \\
Y_{8}^{S}(k,p) &=& -\frac{ q^{2} Y_{2}(k,p) + (k^{2}- p^{2})
Y_{3}(k,p) }{ 8 \nabla(k,p) } \,. \label{Y8 Sym in terms of taus}
\end{eqnarray}

It is also worth noting that the trivial
choice $Y_{i}(k,p)=0$ for all $Y$-functions completely
fixes the transverse vertex, defined through Eqs.~(\ref{transverse
vertex structure},\ref{transverse basis definition},\ref{tau 2
from TTI},\ref{tau 8 from TTI}), in terms of the fermion wave
function renormalization, as reported in ref.~\cite{Qin:2013mta}.
However, we shall show that {\em MR of the electron propagator implies
that these $Y$-functions cannot all be zero simultaneously}.

We can invert relations (\ref{tau 1 from TTI}-\ref{tau 8 from TTI})
to write out the $Y$-functions in terms
of $\tau_i$:
\begin{eqnarray}
Y'_{1}(k,p) &=& - 2 \nabla(k,p) \tau_{1}(k,p) \,,
\label{Y1 in terms of taus} \\
Y'_{2}(k,p) &=& \frac{1}{2} (k^2-p^2) \left[ b(k^2,p^2)-2 \tau_{3}(k,p) \right] \nonumber \\
&+&  t^{2} \, \tau_{6}(k,p)  \,,
\label{Y2 in terms of taus} \\
Y'_{3}(k,p) &=& - \frac{1}{2} q^2 \left[ b(k^2,p^2) - 2 \tau_{3}(k,p) \right] \nonumber \\
&+&  2 \nabla(k,p) \tau_{2}(k,p)
-(k^{2} - p^{2}) \tau_{6}(k,p)  \,, \label{Y3 in terms of taus} \\
Y'_{4}(k,p) &=& \frac{1}{2} \left[ c(k^2,p^2) +  \tau_{5}(k,p) \right] \nonumber \\
&+& \frac{1}{4} \left[ 2 (k^{2}-p^{2}) \tau_{4}(k,p) + t^{2} \tau_{7}(k,p) \right]
\,, \label{Y4 in terms of taus} \\
Y'_{5}(k,p) &=& - \frac{3}{2} q^2 \left[ b(k^2,p^2) - 2 \tau_{3}(k,p) \right] \nonumber \\
&+&  2 \nabla(k,p) \tau_{2}(k,p)
- 3 (k^{2} - p^{2}) \tau_{6}(k,p) \,, \qquad \label{Y5 in terms of taus} \\
{Y'_{6}}^{A}(k,p) &=& -3 \left[ c(k^2,p^2) +  \tau_{5}(k,p) \right] \nonumber \\
&-& \frac{1}{2} \left[ 2 (k^{2}-p^{2}) \tau_{4}(k,p) + t^{2} \tau_{7}(k,p) \right] \,,
\label{Y6 Antisym in terms of taus} \\
{Y'_{7}}^{S}(k,p) &=&  - \left[ 2 q^{2} \tau_{4}(k,p)
+(k^{2}-p^{2}) \tau_{7}(k,p) \right] \,,
 \label{Y7 Sym in terms of taus} \nonumber \\
{Y'_{8}}^{A}(k,p) &=& -\frac{1}{2} \left[
 b(k^2,p^2) + \tau_{8}(k,p) \right] \,. \label{Y8 Antisym in terms of taus}
\end{eqnarray}
Here, we have conveniently defined:
\begin{eqnarray}
  Y_i(k,p) &=& (k^2-p^2) \, Y'_i(k,p) \,.
\end{eqnarray}
We expect the study in terms of $Y_i(k,p)$ to be numerically amicable
as the additional factor of $(k^2-p^2)$ in the numerator eases out
any kinematical singularities in the limit $k^2 \rightarrow p^2$.

%From eqs. (\ref{Y1 in terms of taus}-\ref{Y8 Antisym in terms of
%taus}) it is obvious that the functions $Y_{2}$ and $Y_{7}^{S}$
%are symmetric under $k \leftrightarrow p$, while the rest of the
%$Y$-functions are antisymmetric.

So far, we have shown that the TTIs relate the transverse vertex
form factors to the fermion propagator and a non-local tensor
vertex, but nevertheless this is not enough to elucidate the
analytical behavior of the $Y$-functions. It is
insightful to analyze the asymptotic behavior of the vertex. It
has been shown that in the asymptotic limit, defined as the
perturbative expansion with $p^2 \gg k^2 \gg m^2_0$, the leading
logarithmic term of the transverse vertex reads (in our
kinematical configuration) as~\cite{Curtis:1991fb}:
\begin{eqnarray}
\Gamma_{\mu}^{T} (k,p) \hspace{-.15cm} & \overset{p^2>>k^2}{=} &
\hspace{-.15cm} \frac{\alpha \xi}{8 \pi p^2} \log \left(
\frac{k^2}{p^2} \right) T_{\mu}^{asy} \,, \label{CP a symptotic
vertex}
\end{eqnarray}
where $\xi$ is the gauge-fixing parameter, and
\begin{eqnarray}
T_{\mu}^{asy} \equiv T_{\mu}^{3 \, asy} = T_{\mu}^{6 \, asy} = p^2
\gamma_{\mu} - p_{\mu} \gamma \cdot p \,. \label{aymtotic basis}
\end{eqnarray}
On the other hand, from Eqs.~(\ref{transverse vertex
structure},\ref{transverse basis definition},\ref{tau 1 from
TTI}-\ref{tau 8 from TTI}), it is straightforward to see that the
leading structure of the transverse vertex in the asymptotic limit
acquires the following form:
\begin{eqnarray}
\Gamma_{\mu}^{T} (k,p) \hspace{-.15cm} & \overset{p^2>>k^2}{=} &
\hspace{-.15cm} \left( \tau_3 + \tau_6 \right) T_{\mu}^{asy}
\nonumber \\
& =& \hspace{-.15cm} \frac{\beta}{2 p^2} \log \left(
\frac{k^2}{p^2}
\right) T_{\mu}^{asy} \nonumber \\
& + & \hspace{-.15cm} \Bigg\{ \frac{2 k \cdot q \, Y_2 - k \cdot t
\left( Y_3 - Y_5 \right)}{4 (k^2-p^2) \nabla(k,p)}
\Bigg\} T_{\mu}^{asy} \,, \qquad
\label{Asymptotic vertex in terms of Ys}
\end{eqnarray}
where we have used the fact that the one-loop expansion
of the wave function renormalization yields $F(k^2) = 1 + \beta
\log (k^2/\Lambda^2)$, where $\beta$ is a constant of order
${\cal{O}}(\alpha)$: we shall show in the next section that
$\beta=\alpha \xi/(4\pi)$. Hence, the leading logarithmic
expansion for the asymptotic limit of the vertex, Eq.~(\ref{CP a
symptotic vertex}), demands
%Hence, the leading term of eq. (\ref{Asymptotic vertex in terms of
%Ys}) renders the asymptotic limit of the transverse vertex, eq.
%(\ref{CP a symptotic vertex}), if and only if the following
%relation holds true at least at next to leading order in the
%asymptotic expansion:
\begin{eqnarray}
2 k \cdot q \, Y_2(k,p) = k \cdot t \Big( Y_3(k,p) - Y_5(k,p)
\Big) \,, \label{PT constraint on Y-functions}
\end{eqnarray}
which must be fulfilled at least to second order of its
perturbative expansion in powers of $k^2/p^2$, in order to ensure
the correct asymptotic limit of the transverse vertex.

Although the TTIs, and in particular the identities (\ref{TTI 1})
and (\ref{TTI 2}), are potentially able to fix the transverse
vertex, the construction of an \textit{Ansatz} for this vertex is
far from being complete since the $Y$-functions remain unknown.
Additional requirements need to be implemented in order to compute
them. In this spirit, we shall use the argument of MR
for the fermion propagator, in the chirally symmetric limit, in
order to derive an integral constraint for these $Y$-functions. We
shall also restrict the structure of the vertex by implementing
symmetry arguments and demanding a gauge independent breaking of
chiral symmetry. To this end, we introduce the SDE for the fermion
propagator in the next section.

%%%%%%%%%%%%%%%%%%%%%%%%%%%%%%%%%%%%%%%%%%%%%%%%%%%%%%%%%%%%%%%
%%%%%%%%%%%%%%%%%%%%%%%%%%%%%%%%%%%%%%%%%%%%%%%%%%%%%%%%%%%%%%%
%%%%%%%%%%%%%%%%%%%%     GAP EQUATION      %%%%%%%%%%%%%%%%%%%%
%%%%%%%%%%%%%%%%%%%%%%%%%%%%%%%%%%%%%%%%%%%%%%%%%%%%%%%%%%%%%%%
%%%%%%%%%%%%%%%%%%%%%%%%%%%%%%%%%%%%%%%%%%%%%%%%%%%%%%%%%%%%%%%

\section{Gap equation}
\label{SECTION Gap equation}

\begin{figure}[!ht]
    \centering
    \includegraphics[scale=.5]{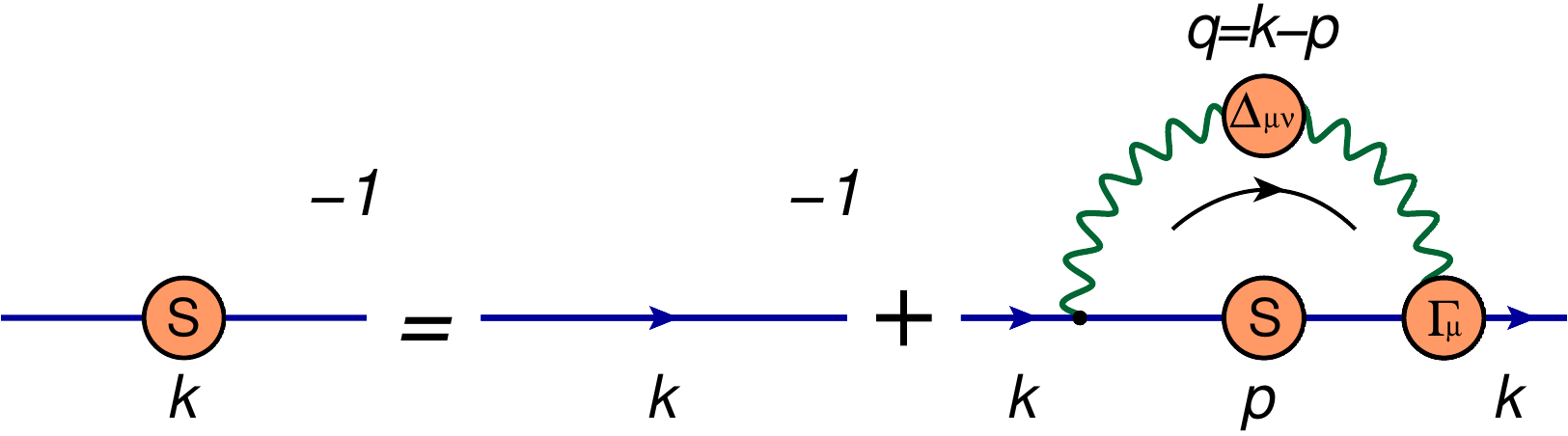}
    \caption{Gap equation for the fermion propagator. The color-filled
    blobs labelled with $S$, $\Delta_{\mu\nu}$ and
    $\Gamma_{\mu}$ stand for the fully-dressed fermion and photon
    propagators, and the three-point vertex, respectively.}
    \label{PropagatorSDEQED}
\end{figure}

The SDE for the fermion propagator, also known as the fermion
\textit{gap equation}, is diagrammatically represented in
Fig.~\ref{PropagatorSDEQED}.

Mathematically, the gap equation is written as:
\begin{eqnarray}
S^{-1}(k) &=& S_{0}^{-1}(k) + \frac{\alpha}{4 \pi^{3}} \int_{E}{
d^{4}p \, \gamma_{\nu} S(p) \Gamma_{\mu}(k,p) \Delta_{\mu\nu} (q)
} \,, \nonumber \\
 \label{gap equation}
\end{eqnarray}
where the subscript $E$ indicates that the integral is performed in the Euclidean space, $\alpha=e^{2}/4 \pi$ is the electromagnetic coupling, and
$\Delta_{\mu\nu} (q)$ is the fully-dressed photon propagator. For
an arbitrary gauge, it is defined as
\begin{eqnarray}
\Delta_{\mu\nu} (q) &=& \Delta(q^2) \left[ \delta_{\mu \nu} -
\frac{q_{\mu} q_{\nu}}{q^{2}} \right] + \, \xi \frac{q_{\mu}
q_{\nu}}{q^{4}} \,, \label{boson propagator definition}
\end{eqnarray}
where $\Delta(q^2)$ is
the photon propagator dressing function. The subscript ``0" in the
first term of the right-hand side of Eq.~(\ref{gap equation})
denotes the tree level fermion propagator.

Recall from Eq.~(\ref{fermion propagator definition}) that the
fermion propagator is defined by the wave function renormalization
and the mass function, so the gap equation, Eq.~(\ref{gap
equation}), can be decomposed into two coupled, integral equations
for ${\cal{M}}$ and $F$, which, in an arbitrary gauge, are
respectively written as:
\begin{eqnarray}
\frac{ {\cal{M}} (k^2) }{ F(k^2) } &=& m_0 + \frac{\alpha \xi}{4
\pi^3} \int_{E} \frac{d^4 p}{q^4} \frac{F(p^2)}{ p^2 + {\cal{M}}^2
(p^2) } \frac{1}{F(k^2)} \nonumber \\
&& \hspace{2cm} \times \Big\{ {\cal{M}} (p^2) \, q \cdot k -
{\cal{M}} (k^2) \, q \cdot p \Big\} \nonumber \\
&+& \frac{\alpha }{4 \pi^3} \int_{E} d^4 p \, \frac{F(p^2)}{ p^2 +
{\cal{M}}^2 (p^2) } {\cal{M}} (p^2) \,
G_{{\cal{M}}} (k,p) \,, \nonumber \\
&& \label{ProyM} \\
\frac{ 1 }{ F(k^2) } &=& 1 - \frac{\alpha \xi}{4 \pi^3} \int_{E}
\frac{d^4 p}{q^4} \frac{F(p^2)}{ p^2 + {\cal{M}}^2 (p^2) }
\frac{1}{F(k^2)} \nonumber \\
&& \hspace{2cm} \times \Big\{ \, q \cdot p + {\cal{M}} (k^2)
{\cal{M}} (p^2) \, \frac{q \cdot k}{k^2} \Big\} \nonumber \\
&+& \frac{\alpha }{4 \pi^3} \int_{E} \frac{d^4 p}{k^2}
\frac{F(p^2)}{ p^2 + {\cal{M}}^2 (p^2) } u(k,p) \, G_F
(k,p) \,, \nonumber \\
&& \label{ProyF}
\end{eqnarray}
where we have adopted the notation $F(k^{2}) \equiv F(k^{2},\Lambda^{2})$
and conveniently defined
\begin{eqnarray}
u(k,p) \equiv 3 k \cdot p - 2 \frac{\nabla(k,p)}{q^2} \,.
\label{u(k,p)-function}
\end{eqnarray}
The functions $G_{F}$ and $G_{{\cal{M}}}$ in
Eqs.~(\ref{ProyM},\ref{ProyF}) encode the effective contribution of the
fully-dressed fermion-photon vertex to the corresponding
equations, and they are defined as
\begin{eqnarray}
&& \hspace{-0.45cm} \frac{(k^2-p^2)}{ \Delta(q^2) } G_{{\cal{M}}} (k,p) = Y_5 (k,p)
+ \frac{\Lambda_M
(p,k)}{ {\cal{M}} (p^2)  } \nonumber \\
&&  \hspace{-0.3cm} + \left\{ 3 k^2 - u(k,p) + \Big[ u(k,p) - 3 p^2
\Big] \frac{ {\cal{M}} (k^2) }{ {\cal{M}} (p^2) }  \right\} \frac{
1 }{ F(k^2) } \,, \qquad  \label{GM(k,p)} \\
&& \hspace{-0.45cm} \frac{(k^2-p^2)}{ \Delta(q^2) } u(k,p) G_F (k,p)= \nonumber \\
&&  -  \Lambda_{NM} (k,p) -
{\cal{M}} (p^2) \Lambda_M (k,p)  \nonumber \\
&& + (k^2-p^2) \left\{ - 3 k^2 p^2 b(k^2,p^2) + u(k,p) \tilde{b}(k^2,p^2) \right. \nonumber \\
&& \hspace{2cm} \left. + {\cal{M}} (p^2) \left(  u(k,p) - 3 k^2 \right)
c(k^2,p^2) \right\} \,, \label{GF(k,p)}
\end{eqnarray}
where
\begin{eqnarray}
  \tilde{b}(k^2,p^2) &=& \frac{1}{k^2-p^2} \left[
  \frac{k^2}{F(k^2)} - \frac{p^2}{F(p^2)}  \right] \,.
\end{eqnarray}
Moreover, we have defined \textit{massive} ($\Lambda_M$) and
\textit{non-massive} ($\Lambda_{NM}$) functions as
\begin{eqnarray}
\Lambda_{M} (k,p) \hspace{-.1cm} &=& \hspace{-.1cm} \frac{1}{2}
Y_1 (k,p) +
q \cdot k \, Y_6^A (k,p) + t \cdot k \, Y_7^S (k,p) \,, \nonumber \\
&&  \label{LambdaM} \\
\Lambda_{NM} (k,p) \hspace{-.1cm} &=& \hspace{-.1cm} \frac{1}{2}
(k^2-p^2) Y_2 (k,p) +
\frac{1}{2} t^2 Y_3 (k,p) \nonumber \\
&& - k \cdot p \, Y_5 (k,p) + 4 \nabla(k,p) Y_8^A (k,p) \,.
 \label{LambdaNM}
\end{eqnarray}

%It is important to note that the symmetry properties of the
%involved $Y$-functions , eqs. (\ref{Y-functions symmetric
%properties},\ref{Y-functions antisymmetric properties}),
%%entail
%\begin{eqnarray}
%\Lambda_{NM} (k,p) = - \Lambda_{NM} (p,k) \,, \label{LambdaNM
%antisymmetric property}
%\end{eqnarray}
%while the massive function $\Lambda_{M}(k,p)$ remains asymmetric.
The reason for referring to $\Lambda_M$ and $\Lambda_{NM}$ as
massive and non-massive functions, respectively, is the following:
from Eqs.~(\ref{ProyM},\ref{GM(k,p)}) it is straightforward to see
that in the chiral limit, where $m_0=0$, a massless solution
(${\cal{M}}=0$) is trivially achieved if $\Lambda_M$ has the mass
function as a global factor, i.e. $\Lambda_M \sim {\cal{M}}$. On
the other hand, the function $\Lambda_{NM}$ does not follow this
argument, and therefore it does not have a dependence
on ${\cal{M}}$ as a global factor. In the same spirit, we shall
refer to $Y_1$, $Y_6^A$ and $Y_7^S$ as massive functions and
$Y_2$, $Y_3$, $Y_5$ and $Y_8^A$ as massless functions.

Note that the contribution of $Y_4$ in
Eqs.~(\ref{ProyM},\ref{ProyF}) cancels out. This is an indication that
the vertex cannot be completely extracted solely from the fermion
propagator SDE. Nonetheless, $Y_4$ can be modelled by relying on
additional information, e.g., the expected anomalous
electromagnetic moment for the corresponding fermion.

\subsection{Quenched QED}
The system of Eqs.~(\ref{ProyM},\ref{ProyF}) has long been studied
using different models for the photon propagator, see
Ref.~\cite{Kizilersu:2014ela} and references therein. For the sake
of simplicity, we limit this work to the well-known quenched
approximation (qQED), where fermion loop contributions to the
photon SDE are neglected and the coupling does not run, which in
turn yields
\begin{eqnarray}
\Delta(q^2) \equiv \Delta_{0}(q^2)=1/q^2 \,. \label{boson
propagator in qQED}
\end{eqnarray}
As we mentioned before, one of the goals of the present
article is to study the impact of the transverse vertex on the DCSB and vice versa.
In particular, we shall investigate the constraints imposed by demanding a gauge
independent DCSB in Section~\ref{SECTION Gauge independent DCSB}.
For this purpose, from now on, we focus our attention on the chiral
limit ($m_0=0$) since this is the most insightful scenario to
elucidate how QED undergoes a phase transition from perturbative
to non-perturbative dynamics as we increase the electromagnetic coupling
($\alpha$) up to the critical value ($\alpha_c$)
where DCSB is triggered. For $\alpha < \alpha_c$
the only possible solution to Eq.~(\ref{ProyM}) in the chiral
limit is ${\cal{M}} (k^2)=0$, but as $\alpha \rightarrow \alpha_c$
a second non-zero solution bifurcates away from the trivial one. The theoretical
prediction for the critical coupling above which DCSB takes place
can be extracted from Eq.~(\ref{ProyM}) through implementing bifurcation
analysis.

  In the vicinity of the critical coupling $\alpha \sim \alpha_c$, the dynamically
generated fermion mass is rather small in comparison with any other
mass scale. Therefore, quadratic and
higher terms in the mass function can formally be neglected. In this case,
Eq.~(\ref{ProyF}) for $F$ and consequently its solution, reduce to that of a
massless theory. Thus, the survey of the renormalization properties of the fermion propagator
in massless QED, and the corresponding implications on
the fermion-photon vertex, is mandatory.

In the next section we show that for a massless fermion in
quenched QED, the wave function renormalization possesses a power
law behavior, which is multiplicatively renormalizable. We also
derive a non-perturbative constraint on the non-massive
$Y$-functions that ensures a MR solution for $F$.

%%%%%%%%%%%%%%%%%%%%%%%%%%%%%%%%%%%%%%%%%%%%%%%%%%%%%%%%%%%%%%%
%%%%%%%%%%%%%%%%%%%%%%%%%%%%%%%%%%%%%%%%%%%%%%%%%%%%%%%%%%%%%%%
%%%%%%%%%%%%%%%%%%%%     MR CONSTRAINTS    %%%%%%%%%%%%%%%%%%%%
%%%%%%%%%%%%%%%%%%%%%%%%%%%%%%%%%%%%%%%%%%%%%%%%%%%%%%%%%%%%%%%
%%%%%%%%%%%%%%%%%%%%%%%%%%%%%%%%%%%%%%%%%%%%%%%%%%%%%%%%%%%%%%%

\section{MR constraints}
\label{SECTION MR constraints}

It is well-known that in QED the gap equation (\ref{gap equation})
leads to a fermion propagator that is logarithmically divergent.
However, we can
define renormalized propagators by absorbing these divergences
into the renormailzation constants ${\cal Z}_{i}$. For massless QED, this multiplicative
renormalization is accomplished by introducing renormalized fields,
fermion field $\psi_{R} = \mathcal{Z}_{2}^{-1/2} \psi$, photon
field $A_{\mu}^{R}=\mathcal{Z}_{3}^{-1/2} A_{\mu}$, and also renormalized
coupling $e_{R}=\mathcal{Z}_{2} \mathcal{Z}_{3}^{1/2} e /
\mathcal{Z}_{1}$. Thus, the MR of the fermion propagator requires
renormalized $F_{R}$ to be related to unrenormalized $F$
through
\begin{equation}
F_{R}(k^{2},\mu^{2}) = \mathcal{Z}_{2}^{-1}(\mu^{2},\Lambda^{2})
F(k^{2},\Lambda^{2}) \,, \label{Renormalization Wave function}
\end{equation}
where $\mu$ plays the role of an arbitrary renormalization scale.
In order to solve Eq.~(\ref{Renormalization Wave function}),
the functions involved are expanded as perturbative series
containing terms of the form $\alpha^{n} \ln^{n}$ (called leading
logarithmic terms). This is known as the leading log approximation (LLA).
In the LLA, we then have
\begin{eqnarray}
F(k^{2},\Lambda^{2}) &=& 1 + \sum_{n=1}^{\infty} \alpha^{n} A_{n}
\ln ^{n} \left( \frac{k^{2}}{\Lambda^{2}} \right) \,, \label{F unrenormalized expansion} \\
\mathcal{Z}_{2}^{-1}(\mu^{2},\Lambda^{2}) &=&  1 +
\sum_{n=1}^{\infty} \alpha^{n} B_{n} \ln ^{n} \left(
\frac{\mu^{2}}{\Lambda^{2}} \right)
\,, \label{Z function} \\
F_{R}(k^{2},\mu^{2}) &=& 1 + \sum_{n=1}^{\infty} \alpha^{n} C_{n}
\ln ^{n} \left( \frac{k^{2}}{\mu^{2}} \right) \,, \label{F
renormalized expansion}
\end{eqnarray}
where $A_{n}$, $B_{n}$ and $C_{n}$ are unknown coefficients but
can be calculated in perturbation theory to any desired order.
However, MR condition (\ref{Renormalization Wave function})
restricts the coefficients to be interrelated as follows:
\begin{equation}
A_{n}=C_{n}=(-1)^{n}B_{n}=\frac{A_{1}^{n}}{n!}\,,
\label{coefficients}
\end{equation}
so that the functions $F$, $F_{R}$ and $\mathcal{Z}_{2}$ obey a
power law behavior. Then, the infinite order solution of
(\ref{F unrenormalized expansion}) for $F$ can be summed up as follows:
\begin{equation}
F(k^{2},\Lambda^{2}) = \left( \frac{k^{2}}{\Lambda^{2}} \right)
^{\beta} \,, \label{F unrenormalized lead log expansion}
\end{equation}
where we define $\beta= \alpha A_{1}$. This is the LLA. Beyond it,
$\beta$ would have terms of ${\cal O}(\alpha^2)$.
Naturally, PT allows us to evaluate the
anomalous dimension $\beta$ at different orders of approximation.

The one-loop contribution to the fermion propagator can be evaluated  by
taking the tree level expressions for $S(p)$, $\Gamma_{\mu}(k,p)$
and $\Delta_{\mu\nu}(q)$ on the right-hand side of Eq.~(\ref{gap
equation}). In the massless limit, ${\cal{M}}=0$, and the resulting
expression for $F$ is
\begin{eqnarray}
\frac{1}{F(k^{2},\Lambda^{2})} &=& 1 + \frac{\alpha \xi}{4
\pi^{3}} \int_{E}{ \frac{d^{4}p}{p^{2}} \frac{ \left[ p^{2} - k
\cdot p \right]}{q^{4}} } \nonumber \\
&-& \frac{\alpha}{4 \pi^{3}} \int_{E}{
\frac{d^{4}p}{p^{2}} \frac{ \left[ 2 \nabla(k,p) -3 q^{2} (k \cdot
p) \right] }{k^{2} q^{4}} } \,. \quad \label{one loop angular integral
for F}
\end{eqnarray}
Angular integration of Eq.~(\ref{one loop angular integral for F})
leads to
\begin{equation}
\frac{1}{F(k^{2},\Lambda^{2})} = 1 + \frac{\alpha \xi}{4 \pi}
\int_{k^{2}}^{\Lambda^{2}}{ \frac{d p^{2}}{p^{2}}} \,. \label{one
loop radial integral for F }
\end{equation}
Carrying out radial integration in Eq.~(\ref{one loop radial
integral for F }) yields
\begin{equation}
F(k^{2},\Lambda^{2}) = 1 + \frac{\alpha \xi}{4 \pi} \log \left(
\frac{k^{2}}{\Lambda^{2}} \right) \,. \label{one loop F}
\end{equation}
Comparing expression (\ref{one loop F}) with the perturbative
expansion (\ref{F unrenormalized expansion}) to one-loop order, we
see that $A_{1}= \xi/4\pi$. Therefore, PT fixes the ${\cal O}(\alpha)$
anomalous dimension in Eq.~(\ref{F unrenormalized lead log
expansion}) to be
\begin{equation}
\beta = \frac{\alpha \xi}{4\pi} \,.  \label{anomalous dimension}
\end{equation}
The power law behavior of $F$ in Eq.~(\ref{F unrenormalized lead log
expansion}), with $\beta$ given in Eq.~(\ref{anomalous
dimension}), is the solution of
\begin{equation}
\frac{1}{F(k^{2},\Lambda^{2})} = 1 + \frac{\alpha \xi}{4\pi}
\int_{k^{2}}^{\Lambda^{2}}{ \frac{d p^{2}}{p^{2}}
\frac{F(p^{2},\Lambda^{2})}{F(k^{2},\Lambda^{2})} } \,.
\label{integral equation for F}
\end{equation}
Note that Eq.~(\ref{integral equation for F}) is non-perturbative in
nature and serves as a requirement of MR for the wave function
renormalization $F$: any \textit{Ansatz} for the three-point
vertex must guarantee that the wave function renormalization $F$ in
Eq.~(\ref{gap equation}) satisfies Eq.~(\ref{integral equation for F}). We shall now proceed to show
how the requirement of MR for the fermion propagator, embodied in
Eq.~(\ref{integral equation for F}), constrains the massless
$Y$-functions.

In the massless limit, Eq.~(\ref{ProyF}) reduces to
\begin{eqnarray}
&& \hspace{-1.2cm} \frac{1}{F(k^{2})} = 1 + \frac{\alpha \xi}{4 \pi^{3}} \int_{E}{
\frac{d^{4}p}{p^{2}} \frac{F(p^{2})}{F(k^{2})} \frac{ \left[ p^{2} - k \cdot p \right]}{q^{4}} } \nonumber \\
&& \hspace{-.8cm} - \frac{\alpha}{4 \pi^{3}} \frac{1}{k^{2}} \int_{E}{
\frac{d^{4}p}{p^{2}} \frac{F(p^{2}) \Delta (q^{2}) }{k^2-p^2} } \Bigg\{ \Lambda_{NM} (k,p) + \nonumber \\
&& \hspace{-0.5cm} (k^2-p^2) \left[
 3 k^2 p^2 b(k^2,p^2) - \tilde{b}(k^2,p^2) u(k,p) \right] \Bigg\} \,. \label{gap equation in terms of
Y-functions}
\end{eqnarray}
Angular integration of the last term on the right-hand side of
the above Eq.~(\ref{gap equation in terms of Y-functions})
vanishes in qQED, since
\begin{eqnarray}
\int_{0}^{\pi} d \varphi \, \sin^2 \varphi \, \frac{ u(k,p)}{q^2}
= 0 \,, \label{u(k,p) angular integral}
\end{eqnarray}
where $\varphi$ is the angle between $k$ and $p$. Bearing in mind
the latter result, Eq.~(\ref{u(k,p) angular integral}), it is
straightforward to see from Eq.~(\ref{gap equation in terms of
Y-functions}) that if we set, quite generally,
\begin{eqnarray}
\Lambda_{NM}(k,p) \hspace{-.1cm} &=& \hspace{-.1cm} (p^2-k^2)  \nonumber \\
\hspace{-.1cm} &\times& \hspace{-.1cm} \left[ {\cal{T}} (k^2,p^2) u(k,p) + 3 k^2 p^2 b(k^2,p^2) \right] , \hspace{.5cm} \label{MR condition on Ys}
\end{eqnarray}
${\cal{T}}$ being an \textit{a priori} arbitrary, dimensionless
function of $k^2$ and $p^2$ alone, then Eq.~(\ref{integral equation for F})
is trivially fulfilled in qQED, i.e., a multiplicatively renormalizable solution
for $F(k^2)$ is ensured.

Symmetry properties of the $Y$-functions,
Eqs.~(\ref{Y-functions symmetric properties},\ref{Y-functions
antisymmetric properties}), restrict ${\cal{T}}(k^2,p^2)$ in
Eq.~(\ref{MR condition on Ys}) to be fully symmetric under $k^2
\leftrightarrow p^2$. Furthermore, in order to ensure that in PT
the transverse form factors start at ${\cal{O}}(\alpha)$, the
perturbative expansion for ${\cal{T}}$ is required to begin at the
same order.

The above expression for $\Lambda_{NM}(k,p)$, Eq.~(\ref{MR
condition on Ys}), provides a non-perturbative \textit{Ansatz} for the
corresponding linear combination of the massless $Y$-functions,
see Eq.~(\ref{LambdaNM}). Although Eq.~(\ref{MR condition on
Ys}) does not fix the $Y$-functions individually, we
shall show in Section \ref{SECTION Gauge independent DCSB} that it
suffices (along with additional constraints on the remaining,
relevant $Y$-functions) to investigate DCSB in the fermion
propagator. Moreover, in Eq.~(\ref{MR condition on Ys}), we assume
that the $q^2$-dependence of the $Y$-functions involved is
effectively incorporated by means of the function $u(k,p)$.
However, more realistic {\em Ans\"{a}tze} are expected to possess a more
complex $q^2$-dependence.

For an arbitrary $q^2$-dependence of $\Lambda_{NM}(k,p)$, it seems
impossible to proceed any further in integrating Eq.~(\ref{gap
equation in terms of Y-functions}) because of the unknown
dependence of the $Y$-functions on the angle $\varphi$. To
circumvent this problem, we shall work with ``\textit{effective}"
functions, denoted as $Y_{i}(k^{2},p^{2})$, whose relation
with the ``\textit{real}" ones, $Y_{i}(k,p)$, is defined exactly in
analogy with~\cite{Bashir:2011vg,Bashir:1997qt} as follows:
\begin{eqnarray}
Y_{2}(k^{2},p^{2}) \hspace{-1mm} & = & \hspace{-1mm}
\frac{1}{f_{2}(k^{2},p^{2})} \int_{0}^{\pi}{ \hspace{-1mm} d
\varphi \sin^{2} \varphi \, \frac{Y_{2}(k,p)}{q^{2}}}
\,, \label{Y2 effective in terms of the real} \\
Y_{3}(k^{2},p^{2}) \hspace{-1mm} & = & \hspace{-1mm}
\frac{1}{f_{3}(k^{2},p^{2})} \int_{0}^{\pi}{ \hspace{-1mm} d
\varphi \sin^{2} \varphi \, \frac{t^{2} \, Y_{3}(k,p)}{q^{2}} }
\,, \label{Y3 effective in terms of the real} \\
Y_{5}(k^{2},p^{2}) \hspace{-1mm} & = & \hspace{-1mm}
\frac{1}{f_{5}(k^{2},p^{2})} \int_{0}^{\pi}{ \hspace{-1mm} d
\varphi \sin^{2} \varphi \, \frac{(k \cdot p) \,
Y_{5}(k,p)}{q^{2}} } \,, \label{Y5 effective in terms of the real} \qquad \\
Y_{8}^{A}(k^{2},p^{2}) \hspace{-1mm}& = & \hspace{-1mm}
\frac{1}{f_{8}(k^{2},p^{2})} \int_{0}^{\pi}{ \hspace{-1mm} d
\varphi \sin^{2} \varphi \, \frac{\nabla(k,p) \,
Y_{8}^{A}(k,p)}{q^{2}} } \,, \nonumber \\
&& \label{Y8 Antisym effective in terms of the real}
\end{eqnarray}
where we have defined
\begin{eqnarray}
f_{2}(k^{2},p^{2}) &=& \int_{0}^{\pi} d \varphi \sin^{2} \varphi
\, \frac{1}{q^{2}} \nonumber \\
&=& \frac{\pi}{2} \left[ \frac{1}{p^{2}} \theta(p^{2}-k^{2}) +
\frac{1}{k^{2}} \theta(k^{2}-p^{2}) \right] \,, \qquad \label{f2 function definition} \\
f_{3}(k^{2},p^{2}) &=& \int_{0}^{\pi} d \varphi \sin^{2} \varphi
\, \frac{t^{2}}{q^{2}} \nonumber \\
&=& \frac{\pi}{2} \left[ \left( 1 + 2 \frac{k^{2}}{p^{2}} \right)
\theta(p^{2}-k^{2}) \right. \nonumber \\
&& \hspace{.5cm}  + \left. \left( 1 + 2 \frac{p^{2}}{k^{2}}
\right) \theta(k^{2}-p^{2}) \right] \,, \label{f3 function
definition} \\
f_{5}(k^{2},p^{2}) &=& \int_{0}^{\pi} d \varphi \sin^{2} \varphi
\, \frac{(k \cdot p)}{q^{2}} \nonumber \\
&=& \frac{\pi}{4} \left[ \frac{k^{2}}{p^{2}} \theta(p^{2}-k^{2}) +
\frac{p^{2}}{k^{2}} \theta(k^{2}-p^{2}) \right] \,, \label{f5 function definition} \\
f_{8}(k^{2},p^{2}) &=& \int_{0}^{\pi} d \varphi \sin^{2} \varphi
\, \frac{\nabla(k,p)}{q^{2}} \nonumber \\
&=& -\frac{\pi}{8} \left[ \frac{k^{2}}{p^{2}} \left( k^{2}-3p^{2}
\right) \theta(p^{2}-k^{2}) \right. \nonumber \\
&& \hspace{.5cm}  + \left. \frac{p^{2}}{k^{2}} \left( p^{2}-3k^{2}
\right) \theta(k^{2}-p^{2}) \right] \,, \label{f8 function
definition}
\end{eqnarray}
where $\theta$ is the usual step function:
\begin{eqnarray}
\theta(x-y)= \left\{ \begin{matrix}   1 \qquad \hbox{for} \qquad x
\geq y \,,
\\ 0 \qquad \hbox{for} \qquad x < y \,.
\end{matrix} \right. \label{step function definition}
\end{eqnarray}
Using aforementioned effective functions, angular integration
of Eq.~(\ref{gap equation in terms of Y-functions}) in qQED leads
to
\begin{eqnarray}
\frac{1}{F(k^{2})} &=& 1 + \frac{\alpha \xi}{4 \pi}
\int_{k^{2}}^{\Lambda^{2}}{ \frac{d p^{2}}{p^{2}} \frac{ F(p^{2})
}{ F(k^{2}) } } \nonumber \\
&& \hspace{-1.2cm} - \frac{\alpha}{4 \pi} \int_{0}^{k^{2}}{
\frac{d p^{2}}{k^{2}} F(p^{2}) \bigg\{ 3 p^{2} b(k^{2},p^{2}) }
\nonumber \\
&& \hspace{-.4cm} + \frac{1}{2 k^{2}} Y_{2}(k^{2},p^{2}) +
\frac{1}{2} \frac{ Y_{3}(k^{2},p^{2})}{k^{2} - p^{2}} \left( 1 + 2
\frac{p^{2}}{k^{2}} \right) \nonumber \\
&& \hspace{-.4cm} - \frac{p^{2}}{2 k^{2}} \frac{
Y_{5}(k^{2},p^{2})}{k^{2} - p^{2}} - p^{2} \frac{
Y_{8}^{A}(k^{2},p^{2})}{k^{2} - p^{2}} \left( \frac{p^{2}}{k^{2}}
-3 \right) \bigg\}  \nonumber \\
&& \hspace{-1.2cm} - \frac{\alpha}{4 \pi}
\int_{k^{2}}^{\Lambda^{2}}{ \frac{d p^{2}}{ k^{2}} F(p^{2})
\bigg\{ 3 k^{2} b(k^{2},p^{2}) } \nonumber \\
&& \hspace{-.4cm} + \frac{1}{2 p^{2}} Y_{2}(k^{2},p^{2}) +
\frac{1}{2} \frac{ Y_{3}(k^{2},p^{2})}{k^{2} - p^{2}} \left( 1 + 2
\frac{k^{2}}{p^{2}} \right) \nonumber \\
&& \hspace{-.4cm} - \frac{k^{2}}{2 p^{2}} \frac{
Y_{5}(k^{2},p^{2})}{k^{2} - p^{2}} - k^{2} \frac{
Y_{8}^{A}(k^{2},p^{2})}{k^{2} - p^{2}}
\left( \frac{k^{2}}{p^{2}}- 3 \right) \bigg\} \,. \nonumber \\
\label{Gap equation Angular integrated}
\end{eqnarray}
In order to ensure MR of the fermion propagator, we demand
$F(k^{2})$ on the left-hand side of Eq.~(\ref{Gap equation Angular
integrated}) to satisfy Eq.~(\ref{integral equation for F});
this imposes the following restriction:
\begin{eqnarray}
&& \hspace{-.8cm} \int_{0}^{k^{2}}{ \frac{d p^{2}}{k^{2}} F(p^{2})
\bigg\{ 3 p^{2} b(k^{2},p^{2}) }
\nonumber \\
&& + \frac{1}{2 k^{2}} Y_{2}(k^{2},p^{2}) + \frac{1}{2} \frac{
Y_{3}(k^{2},p^{2})}{k^{2} - p^{2}} \left( 1 + 2
\frac{p^{2}}{k^{2}} \right) \nonumber \\
&& - \frac{p^{2}}{2 k^{2}} \frac{ Y_{5}(k^{2},p^{2})}{k^{2} -
p^{2}} - p^{2} \frac{ Y_{8}^{A}(k^{2},p^{2})}{k^{2} - p^{2}}
\left( \frac{p^{2}}{k^{2}}
-3 \right) \bigg\}  \nonumber \\
&& \hspace{-1.1cm} + \int_{k^{2}}^{\Lambda^{2}}{ \frac{d p^{2}}{
k^{2}} F(p^{2})
\bigg\{ 3 k^{2} b(k^{2},p^{2}) } \nonumber \\
&& + \frac{1}{2 p^{2}} Y_{2}(k^{2},p^{2}) + \frac{1}{2} \frac{
Y_{3}(k^{2},p^{2})}{k^{2} - p^{2}} \left( 1 + 2
\frac{k^{2}}{p^{2}} \right) \nonumber \\
&& - \frac{k^{2}}{2 p^{2}} \frac{ Y_{5}(k^{2},p^{2})}{k^{2} -
p^{2}} - k^{2} \frac{ Y_{8}^{A}(k^{2},p^{2})}{k^{2} - p^{2}}
\left( \frac{k^{2}}{p^{2}}- 3 \right) \bigg\} = 0 \,. \nonumber \\
\label{Gap equation Constrained}
\end{eqnarray}
This requirement encodes the fact that all divergences have
already been absorbed in the MR solution for the wave function
renormalization $F$. As a consequence, there is no necessity of
regularizing Eq.~(\ref{Gap equation Constrained}) and
we can take $\Lambda^{2} \rightarrow \infty$ in the integration
limit. It is convenient to introduce a dimensionless variable
$x$, defined as
\begin{eqnarray}
&& x=\frac{p^{2}}{k^{2}} \quad \forall \quad p^{2} \in
\left[0,k^{2} \right] \,, \label{X variable definition 1} \\
&& x=\frac{k^{2}}{p^{2}} \quad \forall \quad p^{2} \in
 \left[ k^{2} , \infty \right] \,, \label{X variable definition 2}
\end{eqnarray}
so that Eq.~(\ref{Gap equation Constrained}) is now expressed
as
\begin{equation}
\int_{0}^{1}{ dx \, W(x) } = 0 \,, \label{W restriction}
\end{equation}
with
\begin{equation}
W(x) = 6 \frac{r(x)}{x-1} +\left( x^{\beta} + x^{-2} \right) \Big[
h_{1}(x) + h_{2}(x) \Big] \,, \label{W definition}
\end{equation}
where we have defined the function
\begin{equation}
r(x) = x \left( 1-x^{\beta} \right) - x^{-1} \left( 1-x^{-\beta}
\right) \,. \label{R function definition}
\end{equation}
The presence of the anomalous dimension $\beta$ as an exponent in
Eqs.~(\ref{W definition},\ref{R function definition}), is related
to the terms $F(k^{2})/F(p^{2})$ in (\ref{Gap equation
Constrained}), which can be expressed as $(k^{2}/p^{2})^{\beta}$
in the light of Eq.~(\ref{F unrenormalized lead log expansion}).
Furthermore, in Eq.~(\ref{W definition}), we have defined
\begin{eqnarray}
h_{1}(x) &=& \frac{F(k^{2})}{x-1} \frac{1}{k^{2}} H_{1}(k^{2}, x
k^{2}) \,, \label{h1 function definition}
\\
h_{2}(x) &=& \frac{F(k^{2})}{x-1} \frac{x}{k^{2}} H_{2}(k^{2}, x
k^{2}) \,. \label{h2 function definition}
\end{eqnarray}
These are dimensionless functions, satisfying the properties
\begin{eqnarray}
h_{1}(x^{-1}) &=& x^{\beta-1} h_{1}(x) \,, \label{h1 property} \\
h_{2}(x^{-1}) &=& x^{\beta-2} h_{2}(x) \,. \label{h2 property}
\end{eqnarray}
Moreover, in Eqs.~(\ref{h1 function definition},\ref{h2 function
definition}), we have conveniently defined scalar functions
\begin{eqnarray}
H_{1}(k^{2},p^{2}) &=& \left( \frac{p^{2}}{k^{2}} - 1 \right)
\hspace{-0.1cm} Y_{2}(k^{2},p^{2}) - \hspace{-0.1cm} \left( \frac{p^{2}}{k^{2}} + 1 \right)
\hspace{-0.1cm} Y_{3}(k^{2},p^{2}) \nonumber \\
&& - 8 p^{2} Y_{8}^{A}(k^{2},p^{2}) \,, \label{H1(k2,p2) Definition} \\
H_{2}(k^{2},p^{2}) &=& -Y_{3}(k^{2},p^{2}) + Y_{5}(k^{2},p^{2}) \nonumber \\
&& + 2 (k^{2} + p^{2})  Y_{8}^{A}(k^{2},p^{2}) \,.
\label{H2(k2,p2) Definition}
\end{eqnarray}
Employing $x=p^{2}/k^{2}$ in Eq.~(\ref{W definition}), and using
definitions (\ref{h1 function definition},\ref{h2 function
definition},\ref{H1(k2,p2) Definition},\ref{H2(k2,p2)
Definition}), we have
\begin{eqnarray}
W \left( \frac{p^{2}}{k^{2}} \right) &=& \frac{S(k^{2},p^{2})}{
p^{2}-k^{2} } \Bigg\{ \left( 1 -
\frac{k^{2}}{p^{2}} \right) Y_{2}(k^{2},p^{2}) \nonumber \\
&& \hspace{-1cm} -\left( 2 + \frac{k^{2}}{p^{2}} \right)
Y_{3}(k^{2},p^{2}) + Y_{5}(k^{2},p^{2})  \nonumber \\
&& \hspace{-1cm} + 2 (p^{2}-3k^{2}) Y_{8}^{A}(k^{2},p^{2})
\Bigg\} + 6 k^{2} \, \frac{ r\left( {p^{2}}/{k^{2}} \right) }{ p^{2}-k^{2} } \,, \qquad
\label{W(p2/k2)}
\end{eqnarray}
where we have defined
\begin{eqnarray}
S(k^{2},p^{2}) = F(k^{2}) \frac{k^{2}}{p^{2}} + F(p^{2})
\frac{p^{2}}{k^{2}} \,, \label{S function definition}
\end{eqnarray}
which enters the definition of $r(p^{2}/k^{2})$ through
\begin{equation}
r\left( \frac{p^{2}}{k^{2}} \right) = S(k^{2},p^{2}) \left[
\frac{1}{F(p^{2})} - \frac{1}{F(k^{2})} \right] \,. \label{r
function in terms of S(k2,p2)}
\end{equation}
Eqs.~(\ref{W restriction}-\ref{r function in terms of S(k2,p2)})
constitute non-perturbative constraints on the fermion-photon
vertex: for every \textit{Ansatz} for the $Y$-functions, the resulting
function $W$ is restricted to guarantee the integral constraint
(\ref{W restriction}), so that the MR of the fermion propagator is
ensured. To bring out the applicability and scope of the integral
constraint on the massless $Y$-functions, Eq.~(\ref{W
restriction}), we now proceed to analyze an existing, rather general
transverse vertex \textit{Ansatz}, which was constructed in
qQED to implement the requirement of MR for massless
fermion propagator in addition to all other key features of QED
mentioned before,~\cite{Bashir:2012fs}. Different choices of the free
parameters defining this \textit{Ansatz} correspond to numerous vertices
constructed in the past. We will make reference to all these constructions along the way.

%%%%%%%%%%%%%%%%%%%%%%%%%%%%%%%%%%%%%%%%%%%%%%%%%%%%%%%%%%%%%%%
%%%%%%%%%%%%%%%%%%%%%%%%%%%%%%%%%%%%%%%%%%%%%%%%%%%%%%%%%%%%%%%
%%%%%%%%%%%%%%%%%%%%%%%     EXAMPLE     %%%%%%%%%%%%%%%%%%%%%%%
%%%%%%%%%%%%%%%%%%%%%%%%%%%%%%%%%%%%%%%%%%%%%%%%%%%%%%%%%%%%%%%
%%%%%%%%%%%%%%%%%%%%%%%%%%%%%%%%%%%%%%%%%%%%%%%%%%%%%%%%%%%%%%%

\section{Examples}
\label{SECTION Example}

The Bashir-Bermudez-Chang-Roberts (BBCR) vertex, ref.~\cite{Bashir:2011dp}, is an
\textit{Ansatz} for the dressed fermion-photon vertex in QED, whose
construction is constrained primarily by two requirements: to provide MR
of the fermion propagator and to produce gauge independent
critical coupling for DCSB. As it involves projecting
the vertex onto the gap equation, it is natural that
it is expressed only in terms of the functions which
appear in the full fermion propagator, namely $F(k^{2})$ and
${\cal{M}}(k^{2})$. Moreover, its simplicity lies in the fact that its
functional dependence on these entities is solely through the forms
which enter the longitudinal vertex, namely, $b(k^2,p^2)$ and $c(k^2,p^2)$.
In our kinematical configuration and notation,
the transverse form factors for the BBCR vertex read as
\begin{eqnarray}
\tau_{1}(k^{2},p^{2}) &=& \frac{a_{1}}{(k^{2}+p^{2})} \; c(k^2,p^2) \,, \label{Rocio ansatz tau1} \\
\tau_{2}(k^{2},p^{2}) &=& \frac{a_{2}}{(k^{2}+p^{2})} \; b(k^2,p^2) \,, \label{Rocio ansatz tau2} \\
\tau_{3}(k^{2},p^{2}) &=& a_{3} \; b(k^2,p^2) \,, \label{Rocio ansatz tau3} \\
\tau_{4}(k^{2},p^{2}) &=& \frac{a_{4} (k^2-p^2) }{4 k^2 p^2} \; c(k^2,p^2) \,, \label{Rocio ansatz tau4} \\
\tau_{5}(k^{2},p^{2}) &=& -a_{5} \; c(k^2,p^2) \,, \label{Rocio ansatz tau5} \\
\tau_{6}(k^{2},p^{2}) &=& - \frac{a_{6} (k^2+p^2)}{(k^2-p^2)} \; b(k^2,p^2)
\,, \label{Rocio ansatz tau6} \\
\tau_{7}(k^{2},p^{2}) &=& -\left[  \frac{ a_{4} q^{2} }{ 2 k^{2}
p^{2} } +  \frac{a_{7}}{ k^{2} + p^{2} } \right] \; c(k^2,p^2) \,, \label{Rocio ansatz tau7} \\
\tau_{8}(k^{2},p^{2}) &=& a_{8} \; b(k^2,p^2) \,, \label{Rocio ansatz tau8}
\end{eqnarray}
where the coefficients $a_{i}$ are constants. We will consider this example in
detail because different choices of $a_i$ correspond to several vertices proposed
in the literature, see Ref.~\cite{Aslam:2015nia}, e.g., the Ball-Chiu vertex,~\cite{Ball:1980ay},
the Curtis-Pennington vertex~\cite{Curtis:1990zs} and the Qin-Chang vertex~\cite{Qin:2013mta}.

From Eqs.~(\ref{Y1 in terms of taus}-\ref{Y8 Antisym in terms of
taus},\ref{Rocio ansatz tau1}-\ref{Rocio ansatz tau8}), we see
that the corresponding $Y$-functions for the BBCR vertex read as
\begin{eqnarray}
Y'_{1}(k,p) &=& -2 a_{1} \, c(k^2,p^2) \frac{ \nabla(k,p) }{ k^{2} + p^{2} } \,, \label{Y1 for Rocio vertex} \\
Y'_{2}(k,p) &=& -b(k^2,p^2) \times \nonumber \\
&& \hspace{-1.6cm} \bigg[ (k^{2}-p^{2}) \left( a_{3} - \frac{1}{2} \right) +  a_{6} \left( \frac{ k^{2} + p^{2} }{k^{2} - p^{2}} \right) t^{2} \bigg] , \label{Y2 for Rocio vertex} \\
Y'_{3}(k,p) &=& b(k^2,p^2) \times \nonumber \\
&& \hspace{-1.6cm} \bigg[ q^{2} \left( a_{3} - \frac{1}{2} \right) + 2 a_{2} \frac{ \nabla(k,p) }{ k^{2} + p^{2} } + a_{6} (k^{2} + p^{2}) \bigg] , \label{Y3 for Rocio vertex} \\
Y'_{4}(k,p) &=& - \frac{1}{2} c(k^2,p^2) \times \nonumber \\
&& \hspace{-1.6cm} \bigg[ a_{4}  \frac{\nabla(k,p)}{k^2 p^2} + (a_{5} -1)  + \frac{a_{7}}{2} \frac{t^{2}}{k^2+p^2} \bigg] \,, \label{Y4 for Rocio vertex} \\
Y'_{5}(k,p) &=& 3 b(k^2,p^2) \times \nonumber\\
&& \hspace{-1.6cm} \bigg[ q^{2} \left( a_{3} - \frac{1}{2} \right) + \frac{2}{3} a_{2} \frac{ \nabla(k,p) }{k^{2} + p^{2} } + a_{6} (k^{2} + p^{2}) \bigg] , \label{Y5 for Rocio vertex} \\
{Y'_{6}}^{A}(k,p) &=& c(k^2,p^2) \times \nonumber \\
&& \hspace{-1.6cm} \bigg[ a_{4} \frac{\nabla(k,p)}{k^2 p^2} + 3 (a_{5} -1) + \frac{a_{7}}{2} \frac{t^{2}}{k^2+p^2} \bigg] \,, \label{Y6 for Rocio vertex} \\
{Y'_{7}}^{S}(k,p)&=& a_{7} \, c(k^2,p^2) \frac{ k^{2} -p^{2} }{ k^{2} + p^{2} } \,, \label{Y7 for Rocio vertex} \\
\hspace{-2cm}{Y'_{8}}^{A}(k,p) &=& - \frac{1}{2} b(k^2,p^2) (a_{8} + 1) \,. \hspace{2.4cm} \label{Y8 for Rocio vertex}
\end{eqnarray}
It is mathematically straightforward to show that the asymptotic expansion of
Eqs.~(\ref{Y2 for Rocio vertex},\ref{Y3 for Rocio vertex},\ref{Y5
for Rocio vertex}) in powers of $k^2/p^2$, for $p^2>>k^2>>m_0^2$,
fulfills the PT requirement (\ref{PT constraint on Y-functions})
up to second order if~\cite{Bashir:2011dp}
\begin{eqnarray}
a_3 + a_6 = \frac{1}{2} \,.
\end{eqnarray}

In order to verify that if, for the massless case, the BBCR vertex
satisfies the integral constraint for $W$, Eq.~(\ref{W
restriction}), it is necessary to compute the corresponding
massless effective $Y$-functions by means of Eqs.~(\ref{Y2
effective in terms of the real}-\ref{Y8 Antisym effective in terms
of the real}), which in turn yield
\begin{eqnarray}
Y_{2}(k^2,p^2) &=& b(k^2,p^2) \, \times \nonumber \\
&& \hspace{-1.9cm} \left\{ (1/2-a_{3}) (k^{2}-p^{2})^2
- a_{6} ( k^{2} + p^{2}) (k^2+2 p^2)  \right\}  \,, \qquad
\label{Effective Y2 for Rocio Vertex} \\
Y_{3}(k^{2},p^{2}) &=& \frac{1}{2} \frac{k^4-p^4}{k^2+2 p^2} \; b(k^2,p^2) \times \nonumber \\
&&  \hspace{-1.9cm} \Big\{ (2 a_{3} + 2 a_{6} - 1) k^{2}
-(a_{2} - 4a_{6}) p^{2} \nonumber \\
&& \hspace{-1.2cm} + a_{2} p^2 \left( ({4 k^{4} + 6 k^{2} p^{2} -p^{4}})/({(k^{2} + p^{2})^2})
\right)  \Big\} \,, \label{Effective Y3 for Rocio Vertex} \\
Y_{5}(k^{2},p^{2}) &=& \frac{k^2-p^2}{k^2+p^2} \; b(k^2,p^2) \times \nonumber \\
&& \hspace{-1.9cm} \left\{ 3 a_{6} (k^{2} + p^{2})^2 +  a_{2} p^2 ( k^{2} - p^{2}/2) \right\} \,, \label{Effective Y5 for Rocio Vertex} \\
Y_{8}^{A}(k^{2},p^{2}) &=& - \frac{1}{2} (a_{8} +1) (k^2 - p^2) \; b(k^2,p^2) \,. \label{Effective Y8
for Rocio Vertex}
\end{eqnarray}
Using Eqs.~(\ref{Effective Y2 for Rocio Vertex}-\ref{Effective Y8
for Rocio Vertex}) in the expression for $W(p^{2}/k^{2})$,
Eq.~(\ref{W(p2/k2)}), we find
\begin{eqnarray}
W(x) &=& -2 (a_{2} +2a_{3} -2a_{8}) \frac{r(x)}{1-x} \nonumber \\
&& \hspace{-1.5cm} + \frac{1}{2} \Big[ -3(1 -2a_{6}) + (a_{2} + 2a_{3} -2a_{8}) \Big] \left( \frac{1
+x}{1 -x} \right) r(x) \,. \nonumber \\
\label{W(x) for Rocio Vertex}
\end{eqnarray}
If we insert the above expression for $W(x)$, Eq.~(\ref{W(x) for
Rocio Vertex}), in its corresponding integral restriction,
Eq.~(\ref{W restriction}), we find that the integral vanishes if and
only if the following two conditions are met:
\begin{eqnarray}
a_{2} + 2a_{3} -2a_{8} &=&0 \,, \\
a_{6} &=& {1}/{2} \,,
\label{Rocio constraints on a-coeficients}
\end{eqnarray}
which are the constraints reported in~\cite{Bashir:2011dp} for the
coefficients $a_{i}$ in order to ensure the MR of the fermion
propagator.

As a counterexample, we could take all $Y$-functions equal to
zero. If we do so, the resulting function $W(x)$ reads as
\begin{eqnarray}
W(x)=-6 \frac{r(x)}{1-x} \,, \label{W(x) for Craig Vertex}
\end{eqnarray}
which does not satisfy the integral constraint (\ref{W
restriction}). Therefore, {\em setting all $Y$-functions equal to zero
does not ensure the MR of the fermion propagator}.

Throughout Sections \ref{SECTION MR constraints} and \ref{SECTION
Example}, we have investigated MR
solution for massless fermion propagator in qQED, within the
LLA, and derived a consequent non-perturbative, integral
constraint for the massless $Y$-functions. In the next section, we
study DCSB and implement
the argument of a gauge independent critical coupling to impose
further constraints on the transverse vertex.

%%%%%%%%%%%%%%%%%%%%%%%%%%%%%%%%%%%%%%%%%%%%%%%%%%%%%%%%%%%%%%%
%%%%%%%%%%%%%%%%%%%%%%%%%%%%%%%%%%%%%%%%%%%%%%%%%%%%%%%%%%%%%%%
%%%%%%%%%%%%%%%     GAUGE INDEPENDENT DCSB     %%%%%%%%%%%%%%%%
%%%%%%%%%%%%%%%%%%%%%%%%%%%%%%%%%%%%%%%%%%%%%%%%%%%%%%%%%%%%%%%
%%%%%%%%%%%%%%%%%%%%%%%%%%%%%%%%%%%%%%%%%%%%%%%%%%%%%%%%%%%%%%%

\section{Gauge independent DCSB}
\label{SECTION Gauge independent DCSB}

In order to study DCSB
through the gap equation by employing fully-dressed
fermion-photon vertex, Eqs.~(\ref{Ball-Chiu vertex
decomposition}-\ref{longitudinal coefficients
definitions},\ref{transverse vertex structure},\ref{transverse
basis definition},\ref{tau 1 from TTI}-\ref{tau 8 from TTI}), we
propose an \textit{Ansatz} for the functions $Y_i$ appearing in
Eqs.~(\ref{ProyM}-\ref{LambdaNM}). Naturally, we look for the simplest construction which incorporates all the key constraints we have enlisted and studied so far. This can be achieved by requiring the following~:
\begin{itemize}
\item \textbf{1)} The massive $Y$-functions in the gap equation
are expressed solely in terms of the fermion dressing functions,
$F$ and ${\cal{M}}$.

\item \textbf{2)} The antisymmetric contribution of $G_{\cal{M}}
(k,p)$ and $G_F (k,p)$ vanishes under $k \leftrightarrow p$.

\item \textbf{3)} The functions $G_{\cal{M}} $ and $G_F$ are the
same (up to a constant factor).

\end{itemize}
These simplifying requirements do not jeopardize the MR of the fermion propagator which
can still be ensured in massless qQED.
%In order to study DCSB in the gap equation using the fully-dressed
%fermion-photon vertex, eqs. (\ref{Ball-Chiu vertex
%decomposition}-\ref{longitudinal coefficients
%definitions},\ref{transverse vertex structure},\ref{transverse
%basis definition},\ref{tau 1 from TTI}-\ref{tau 8 from TTI}), we
%conveniently fix the following massive $Y$-functions as

Assumptions~\textbf{1)} and~\textbf{2)} are fulfilled
if we choose
\begin{eqnarray}
Y_1 (k,p) &=& - 4 \frac{\nabla(k,p) (k^2-p^2))}{q^2} \, c(k^2,p^2)
\,, \label{Ansatz for Y1} \\
Y_6^A (k,p) &=& -3 (k^2-p^2) \, c(k^2,p^2)  \,, \label{Ansatz for Y6A} \\
Y_7^S (k,p) &=& 0 \,. \label{Ansatz for Y7S}
\end{eqnarray}
%which leads to the fully symmetrization of the functions
%$G_{\cal{M}}$ and $G_F$ under $k \leftrightarrow p$. For the
%massless $Y$-functions, we demand the MR condition, eq. (\ref{MR
%condition on Ys}), and fix $Y_5$ as
Moreover, we can implement the simplifying requirement~\textbf{3)}
by demanding the MR condition on the massless
functions, Eq.~(\ref{MR condition on Ys}), and fixing $Y_5$ as follows:
\begin{eqnarray}
Y_5 (k,p) = (k^2-p^2) \left[  3 {\cal{T}} (k^2,p^2)  +
u(k,p) b(k^2,p^2) \right] \hspace{-.1cm} . \quad
\label{Ansatz for Y5}
\end{eqnarray}
In fact, the \textit{Ansatz} for the $Y$-functions, constituted through Eqs.~(\ref{MR condition on Ys},\ref{Ansatz for Y1}-\ref{Ansatz for Y5}),
yields
\begin{eqnarray}
\frac{1}{3} G_{{\cal{M}}}(k,p) = G_F(k,p) \hspace{-1.5mm} &=& \hspace{-1.5mm}
 \Delta (q^2) \left[ {\cal{T}} (k^2,p^2)
+ \tilde{b}(k^2,p^2) \right]  \nonumber \\
&\equiv& \hspace{-1.5mm} G(k,p) \,, \label{G-function}
\end{eqnarray}
which fulfills assumption \textbf{2)} in addition to \textbf{3)}. Moreover, for
the massless limit in qQED, it simplifies Eq.~(\ref{ProyF}) as
\begin{eqnarray}
\frac{ 1 }{ F(k^2) } &=& 1 - \frac{\alpha \xi}{4 \pi^3} \int_{E}
\frac{d^4 p}{p^2} \frac{F(p^2)}{ F(k^2) } \frac{q \cdot p}{q^4}
\,. \label{F using our ansatz}
\end{eqnarray}
After angular integration, $F$ satisfies Eq.~(\ref{integral equation for F}) as expected, i.e., it has the power law behavior
of Eq.~(\ref{F unrenormalized lead log expansion}) as constrained
by MR, with the anomalous dimension $\beta$ given in
Eq.~(\ref{anomalous dimension}).

Let us summarize below the important characteristics of the function ${\cal{T}}$:
\begin{itemize}
\item \textbf{i)} It must be a dimensionless function of $k^2$ and $p^2$. We assume it to be $q^2$-independent in order to ensure the MR of $F(k^2)$.

\item \textbf{ii)} It must be fully symmetric under $k^2 \leftrightarrow p^2$.

\item \textbf{iii)} Its perturbative expansion must start at ${\cal O}(\alpha)$.

\item \textbf{iv)} It must vanish in the Landau gauge, $\xi=0$.
\end{itemize}
Recall from Section \ref{SECTION MR constraints} that condition
\textbf{i)} is required to ensure a MR solution for the massless
fermion propagator in qQED, while the conditions \textbf{ii)} and
\textbf{iii)} follow from the symmetry properties of the vertex and
its perturbative expansion, respectively. The additional condition
\textbf{iv)} is imposed in order to facilitate the extraction of an
$\xi$-independent critical coupling, and an anomalous dimension
for the mass function which, at criticality, is independent of the
choice of the vertex, as we shall discuss now.

In the vicinity of the critical coupling, $\alpha_c$, above which chiral symmetry is broken dynamically, the generated fermion
mass is negligible in comparison with any other mass scale. Hence, for
$\alpha \sim \alpha_c$, we can formally neglect quadratic and higher powers (if any) of the mass
function in Eq.~(\ref{ProyM}). In the limit $m_0=0$, it
reduces to
\begin{eqnarray}
\frac{ {\cal{M}} (k^2) }{ F(k^2) } &=&  \frac{\alpha \xi}{4 \pi^3}
\int_{E} \frac{d^4 p}{p^2} \frac{F(p^2)}{ F(k^2) } \frac{1}{q^4} \nonumber \\
&& \hspace{1cm} \times \Big\{ {\cal{M}} (p^2) \, q \cdot k -
{\cal{M}} (k^2) \, q \cdot p \Big\} \nonumber \\
&+& \frac{3\alpha }{4 \pi^3} \int_{E} \frac{d^4 p}{p^2} \, F(p^2)
{\cal{M}} (p^2) \, G (k,p) \,, \label{ProyM at criticality}
\end{eqnarray}
where our \textit{Ansatz} for the $Y$-functions, Eqs.~(\ref{MR condition on
Ys},\ref{Ansatz for Y1}-\ref{Ansatz for Y5}), has been implicitly
embedded through the function $G(k,p)$, defined in Eq.~(\ref{G-function}). Moreover, neglecting terms quadratic in
${\cal{M}}$, the equation for $F$, Eq.~(\ref{ProyF}), reduces to
that of a massless theory and decouples from that of the mass function ${\cal{M}}$. In the quenched approximation, it
yields Eq.~(\ref{F using our ansatz}). Therefore,
from Eqs.~(\ref{F using our ansatz},\ref{ProyM at criticality}) we see that
in the vicinity of the critical coupling, in qQED, the mass
function satisfies the following equation:
\begin{eqnarray}
{\cal{M}} (k^2)  &=&  \frac{\alpha \xi}{4 \pi^3} \int_{E}
\frac{d^4 p}{p^2} \frac{F(p^2)}{ F(k^2) } {\cal{M}} (p^2)
\frac{q \cdot k}{q^4} \nonumber \\
&+& \frac{3\alpha }{4 \pi^3} \int_{E} \frac{d^4 p}{p^2} \, F(p^2)
{\cal{M}} (p^2) \, G (k,p) \,.  \label{ProyM at criticality in
qQED}
\end{eqnarray}
In the neighborhood of
$\alpha_c$, MR forces a power law behavior for the mass function
which must hold at all momenta:
\begin{eqnarray}
{\cal{M}}(k^2) = B_\Lambda \big( k^2 \big)^{-s} \,, \label{MR
solution for M}
\end{eqnarray}
where $B_\Lambda$ is a constant (that depends on $\Lambda$), and
the exponent $s = 1- \gamma_m/2$ is defined in terms of the
anomalous dimension of the mass function, $\gamma_m$. We assume $0
< s \leq 1$ to comply with perturbation theory.

Bardeen \textit{et al.} demonstrated that, at $\alpha = \alpha_c$,
the mass anomalous dimension is $\gamma_m = 1$~\cite{Bardeen:1985sm,Leung:1985sn}. Some further analyses, based on
Cornwall-Jackiw-Tomboulis effective potential technique, tend to argue
that, at criticality, this value holds true regardless of the choice
of the vertex~\cite{Akram:2012jq,Holdom:1988gs,Holdom:1988gr}. In Ref.~\cite{Bashir:1994az}, this values is quite close to unity, though not exactly equal to it. Setting $\gamma_m=1$ results in the four-fermion interaction operator $(\overline{\psi} \psi)^2$ acquiring dynamical
dimension $d=2(3-\gamma_m)=4$ in contrast to its canonical
dimension $d=6$. Therefore, four-fermion interaction becomes
marginal. It must then be included in order to render non-perturbative QED a self-consistent, closed theory~\cite{Akram:2012jq,Bardeen:1990im}.
Depending upon the non-perturbative details of the
fermion-photon interaction, it is plausible to have $\gamma_m > 1$,
implying $d<4$, which would modify the status of the four-point
operators from marginal to relevant; see, e.g., the
review article~\cite{Bashir:2004mu} and references therein.

At $\alpha=\alpha_c$, the anomalous dimension $\gamma_m=1$ and its
corresponding critical value $s_c=1/2$ can be obtained by
constraining the fermion-photon vertex, a line of action which is
followed in~\cite{Bashir:1995qr,Akram:2012jq}. In our analysis, this critical
value is readily derived from Eq.~(\ref{ProyM at criticality}) in
Landau gauge, $\xi=0$, if we demand ${\cal{T}}(k^2,p^2)$ to
fulfill condition \textbf{iv)}: from Eq.~(\ref{F using our
ansatz}) we see that in the Landau gauge $F(k^2)=1$, and therefore
Eq.~(\ref{ProyM at criticality in qQED}) reduces to
\begin{eqnarray}
{\cal{M}} (k^2) = \frac{3 \alpha}{4 \pi^3} \int_{E} \frac{d^4
p}{p^2} \frac{ {\cal{M}} (p^2) }{q^2} \,. \label{Mass function in
Landau gauge}
\end{eqnarray}
For the MR solution of the mass function, Eq.~(\ref{MR solution
for M}), the above Eq.~(\ref{Mass function in Landau gauge}) results in
\begin{eqnarray}
s = \frac{1}{2} \pm \frac{1}{2} \sqrt{1 - \frac{\alpha}{\alpha_c}}
\,, \label{Mass anomalous dimension}
\end{eqnarray}
where $\alpha_c$ stands for the critical coupling, which signals
the point where the two possible solution for $s$ match each
other, and a non-trivial solution for the mass
function\footnote{In addition to positive-definite solutions for
the mass function (and their corresponding mirror), an arbitrary
vertex may produce spurious oscillatory solutions for ${\cal{M}}$.
However, it has been argued that a realistic vertex might only
produce monotonically decreasing and increasing non-trivial
solutions~\cite{Raya:2013ina}.} bifurcates away from the
perturbative one (${\cal{M}}=0$):
for $\alpha > \alpha_c$, the solution for the mass function enters
the complex plane indicating that DCSB has taken place. In this
case, the critical coupling is:
\begin{eqnarray}
\alpha_c = \frac{\pi}{3} \,, \label{critical coupling}
\end{eqnarray}
thus revealing a Miransky scaling law for the interaction strength
$\alpha$~\cite{Fomin:1976af,Fomin:1978rk,Maskawa:1974vs}, which
has been derived using a bare
vertex~\cite{Miransky:1984ef,Fomin:1984tv}. For
$\alpha=\alpha_c=\pi/3$ in Eq.~(\ref{Mass anomalous dimension}),
the expected critical value for the anomalous mass dimension is
obtained, i.e.
\begin{eqnarray}
s_c = \frac{1}{2} \,. \label{critical mass anomalous dimension}
\end{eqnarray}

Since the critical coupling pinpoints a phase transition from
perturbative to non-perturbative dynamics, it is potentially a
physical observable, and hence it is expected to be independent of
the gauge parameter. Thus, Eq.~(\ref{Mass function in Landau
gauge}), and its corresponding MR solution, Eq.~(\ref{MR solution
for M}), must hold in all gauges. Therefore, from Eqs.~(\ref{ProyM
at criticality in qQED},\ref{Mass function in Landau gauge}), we
see that in order for $\alpha_c$ and $s_c$ to be
$\xi$-independent, the mass function and the vertex must satisfy
the following equation in qQED:
\begin{eqnarray}
&& \hspace{-1cm} \int_{E} \frac{d^4 p}{p^2} \frac{ {\cal{M}}(p^2) F(p^2)
}{q^2} \times \nonumber \\
&& \hspace{-.3cm} \Bigg\{ \frac{\xi q \cdot
k}{3 q^2 F(k^2)}  +
{\cal{T}} (k^2,p^2) + k^2 b(k^2,p^2) \Bigg\} = 0 \,.  \label{Mass Restriction Equation}
\end{eqnarray}
After performing angular integration, and introducing the
dimensionless variables defined in Eqs.~(\ref{X variable
definition 1},\ref{X variable definition 2}), the above Eq.~(\ref{Mass Restriction Equation}) can be cast in the following form:
\begin{eqnarray}
\int_{0}^{1} \frac{dx}{\sqrt{x}} V(x) = 0 \,, \label{MR condition
on T(k2,p2) from critical coupling}
\end{eqnarray}
with
\begin{eqnarray}
&& \hspace{-1.5cm} V(x) = \frac{\xi}{3} x^{\beta-s + \nicefrac{1}{2} }+ \left[
x^{\beta-s + \nicefrac{1}{2}} + x^{s - \nicefrac{1}{2}} \right]
g(x) \nonumber
\\
&& \hspace{-.6cm} + \left( \frac{x^{\beta}-1}{1-x} \right) x^{-s +
\nicefrac{1}{2}} - \left( \frac{x^{-\beta}-1}{1-x} \right) x^{s +
\nicefrac{1}{2} }  \,, \label{V definition}
\end{eqnarray}
where $s$ and $\beta$, defined in Eqs.~(\ref{anomalous
dimension},\ref{critical mass anomalous dimension}), appear in the
light of the MR solutions for $F$ and ${\cal{M}}$, Eqs.~(\ref{F
unrenormalized lead log expansion},\ref{MR solution for M}),
respectively. Furthermore, in Eq.~(\ref{V definition}) we have
defined
\begin{eqnarray}
g(x) = F(k^2) {\cal{T}}(k^2,x k^2) \,, \label{g function
definition}
\end{eqnarray}
which is independent of $k^2$, and satisfies the
following property:
\begin{eqnarray}
g(x^{-1}) = x^{\beta} g(x) \,. \label{g property}
\end{eqnarray}

It is important to stress the fact that Eq.~(\ref{MR condition on T(k2,p2) from critical coupling}) stands for a non-perturbative constraint on the vertex: any \textit{Ansatz} for ${\cal{T}}(k^2,p^2)$ must provide a function $V$ that should satisfy Eq.~(\ref{MR condition on T(k2,p2) from critical coupling}). Conversely, from a particular solution for $V$ in the latter equation, one can derive the corresponding function ${\cal{T}}$ by means of Eq.~(\ref{V definition}). However, there exist an infinite number of solutions for $V(x)$ satisfying Eq.~(\ref{MR condition on T(k2,p2) from critical coupling}). In addition, such a solution for $V$ must also satisfy (\textit{cf.} Eq.~(\ref{V definition}))
\begin{eqnarray}
V(x)-V(x^{-1}) = \frac{\xi}{3} \left( x^{\beta-s+\nicefrac{1}{2}} - x^{-\beta+s-\nicefrac{1}{2}}  \right) \,. \label{Condition on V(x)}
\end{eqnarray}
A simple choice satisfying Eqs.~(\ref{MR condition on T(k2,p2) from critical coupling},\ref{Condition on V(x)}) at criticality reads:
\begin{eqnarray}
V(x) = \frac{\xi}{3} \left\{ x^{\beta} + \frac{1 - 2 \beta}{8 \beta^2} \left( 2 - x^{\beta} - x^{-\beta} \right) \right\} \,, \label{ansatz for V(x)}
\end{eqnarray}
valid for $-\nicefrac{1}{2} \leq \beta \leq \nicefrac{1}{2}$ but $\beta \neq 0 $. In the Landau gauge, Eqs.~(\ref{MR condition on T(k2,p2) from critical coupling},\ref{Condition on V(x)}) are satisfied with the trivial solution
\begin{eqnarray}
V(x)_{\xi=0} = 0 \,. \label{ansatz for V(x) Ladau Gauge}
\end{eqnarray}
It is worth reminding that Eq.~(\ref{MR condition on T(k2,p2) from critical coupling}), and consequently Eqs.~(\ref{V definition},\ref{Condition on V(x)}), are rigourously valid only at criticality. Therefore, for $\alpha_c=\pi/3$ and $s_c=1/2$, the resulting function ${\cal{T}}(k^2,p^2)$, derived form Eqs.~(\ref{V definition},\ref{ansatz for V(x)}), reads (for $x=p^2/k^2$ and $-6 \leq \xi \leq 6$, but $\xi \neq 0$) as
\begin{eqnarray}
{\cal{T}} (k^2,p^2) &=& -\frac{1}{2} \left( k^2+p^2 \right) b \left( k^2,p^2 \right) \nonumber \\
&& \hspace{-1cm} + \frac{\left(12-\xi\right)}{2\xi} \left[ \frac{F(k^2)-F(p^2)}{F(k^2)+F(p^2)} \right] \left[ \frac{1}{F(k^2)} - \frac{1}{F(p^2)} \right] \,,
\nonumber \\
\label{ansatz for T(k2,p2)}
\end{eqnarray}
whereas Eqs.~(\ref{V definition},\ref{ansatz for V(x) Ladau Gauge}) yield (for Landau gauge)
\begin{eqnarray}
{\cal{T}} (k^2,p^2)_{\xi=0} = -\frac{1}{2} \left( k^2+p^2 \right) b \left( k^2,p^2 \right) \,.
\label{ansatz for T(k2,p2) Landau Gauge}
\end{eqnarray}
The above expressions for ${\cal{T}}$ fulfill conditions
\textbf{i)}-\textbf{iv)}, as expected, but a few observations must be made: \textbf{a)} MR of the wave function renormalization entails $\nicefrac{F(p^2,\Lambda^2)}{F(k^2,\Lambda^2)} = \nicefrac{F(p^2,\mu^2)}{F(k^2,\mu^2)}$ for some renormalization scale, $\mu^2$, ensuring the $\Lambda^2$-independence of the second term on the RHS of Eq.~(\ref{ansatz for T(k2,p2)}); and \textbf{b)} although $F=1$ for $\xi=0$ (in the LLA of qQED), leading to ${\cal{T}}_{\xi=0}=0$, the function ${\cal{T}}_{\xi=0}(k^2,p^2)$ defined in Eq.~(\ref{ansatz for T(k2,p2) Landau Gauge}) does not necessarily vanish in the Landau gauge beyond the LLA and the quenched approximation.

\begin{figure}[!ht]
    \centering
    \includegraphics[scale=.33]{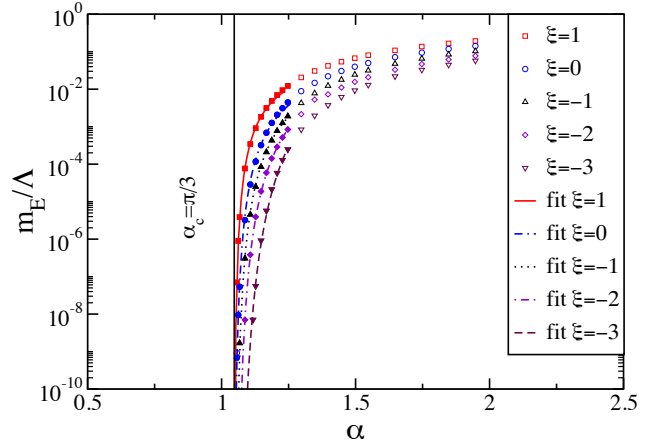}
    \caption{Ratio of the Euclidean mass and the ultraviolet cut-off in different gauges. $\alpha_c=\pi/3$ within the numerical accuracy of our computation.}
    \label{Euclidian Mass Arbitrary Gauge}
\end{figure}

Numerical evaluation of the ratio between the euclidean mass $m_E$, defined as $m_{E} = {\cal{M}}(m_{E}^{2})$, and the ultraviolet cut-off $\Lambda$ is shown in Fig.~\ref{Euclidian Mass Arbitrary Gauge} for different gauges implementing the {\em Ans\"{a}tze} introduced in Eqs.~(\ref{ansatz for T(k2,p2)},\ref{ansatz for T(k2,p2) Landau Gauge}). In the plot, points with $\alpha < 1.25$ (highlighted as filled markers) are fitted to the Miransky scaling law~\cite{Fomin:1976af,Fomin:1978rk,Maskawa:1974vs,Miransky:1984ef,Fomin:1984tv}
\begin{eqnarray}
\frac{m_E}{\Lambda} = \exp \left[ -\frac{ \pi \kappa }{ \sqrt{ \frac{\alpha}{\alpha_c} -1 }} + \phi \right] \,.
\label{Miransky Scaling Law}
\end{eqnarray}
\begin{table}[htp]
 %%%%%%%%%%%%%%%%%%%%%%%%%%%%%%%%%%%%%%%%%%%%%%%%%%%
\begin{center}
\begin{tabular}[b]{| c | c | c | c |}%
\hline
\hspace{.3cm} $\xi$ \hspace{.3cm} & \hspace{.1cm}$\alpha_c=\pi/3$ \hspace{.1cm} & \hspace{.5cm} $\kappa$ \hspace{.5cm} & \hspace{.5cm} $\phi$ \hspace{.5cm} \\
\hline
\rule{0ex}{2.0ex}
$1$ & $0.69\%$ & $0.695$ & $0.481$\\
\rule{0ex}{2.0ex}
$0$ & $0.82\%$ & $0.970$ & $1.383$ \\
\rule{0ex}{2.0ex}
$-1$ & $0.51\%$ & $1.112$ & $1.604$\\
\rule{0ex}{2.0ex}
$-2$ & $0.51\%$ & $1.199$ & $1.679$ \\
\rule{0ex}{2.0ex}
$-3$ & $2.81\%$ & $1.283$ & $1.829$ \\
\hline
\end{tabular}
\end{center}
\caption{\label{table-M}
Parameters for different gauges}
\label{Parametrization Arbitrary Gauges}
\end{table}

Each fit yields the critical coupling $\alpha_c=\pi/3$ with a numerical error of less than 1\% for $\xi=1,0,-1,-2$ (and $\sim$ 2.8\% for $\xi=-3$) as indicated in the second column of the Table \ref{Parametrization Arbitrary Gauges}. It is worth reminding that our {\em Ansatz} reveals $\alpha_c=\pi/3$, just like the bare vertex in the Landau gauge. However, {\em the bare vertex leads to a highly gauge dependent $\alpha_c$, including no chiral symmetry breaking for $\xi=-3$.} In our case, {\em chiral symmetry is broken in every gauge with the same critical coupling.} The result for $\xi=-3$ particularly emphasizes this point.

It is well-known that DCSB manifests itself in the three-point vertex through its massive form factors~\cite{Braun:2014ata,Mitter:2014wpa}, and thus a physically meaningful {\em Ansatz} for ${\cal{T}}(k^2,p^2)$ should incorporate the mass function. In the present work, we propose a simple, numerically tractable {\em Ansatz} for $\xi=0$,
\begin{eqnarray}
{\cal{T}} (k^2,p^2)_{\xi=0} &=& -\frac{1}{2} \left( k^2+p^2 \right) b \left( k^2,p^2 \right) \nonumber \\
&& \hspace{-1.5cm} + \rho \left[ \frac{{\cal{M}}(k^2)}{F(k^2)} + \frac{{\cal{M}}(p^2)}{F(p^2)} \right] c \left( k^2,p^2 \right) \,,
\label{ansatz for T(k2,p2) With Rho}
\end{eqnarray}
which is an extension of Eq.~(\ref{ansatz for T(k2,p2) Landau Gauge}), with a mass term weighted by a real constant $\rho$. The fact that the last term of Eq.~(\ref{ansatz for T(k2,p2) With Rho}) contains quadratic powers in ${\cal{M}}$ ensures that this contribution can be neglected at criticality, which in turn yields a gauge independent critical coupling. In addition, characteristics \textbf{i)}-\textbf{iv)} remain preserved.

\begin{figure}[!ht]
    \centering
    \includegraphics[scale=.33]{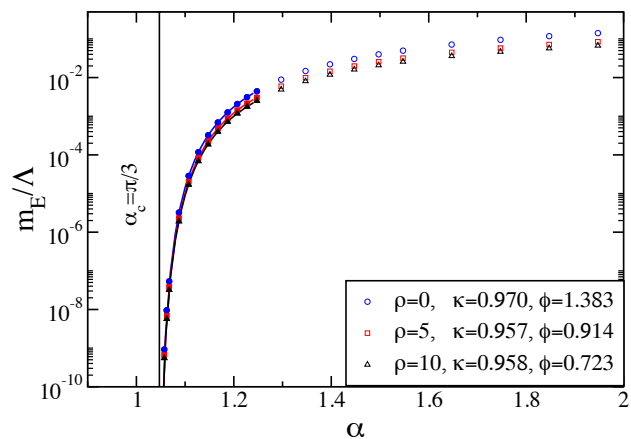}
    \caption{Euclidean mass in Landau gauge for $\rho=0,5,10$. $\rho$ stands for the strength of the DCSB in the vertex,~Eq.~(\ref{ansatz for T(k2,p2) With Rho})}.
    \label{Euclidian Mass Different Rho}
\end{figure}

Numerical evaluation of $m_E/\Lambda$ using Eq.~(\ref{ansatz for T(k2,p2) With Rho}) is shown in Fig.~\ref{Euclidian Mass Different Rho} for different values of $\rho$. Again, points with $\alpha<1.25$ are fitted to Eq.~(\ref{Miransky Scaling Law}),
indicating a critical coupling independent of the $\rho$ parameter and equal to $\pi/3$, within a margin of error smaller than $1\%$.

Numerical results for the Euclidean mass plotted in Figs.~\ref{Euclidian Mass Arbitrary Gauge} and \ref{Euclidian Mass Different Rho} support the argument that any function ${\cal{T}}(k^2,p^2)$ preserving characteristics \textbf{i)}-\textbf{iv)} and satisfying conditions defined through
Eqs.~(\ref{MR condition
on T(k2,p2) from critical coupling},\ref{Condition on V(x)}) will ensure a gauge independent critical coupling $\alpha_c=\pi/3$ in qQED.
Eqs.~(\ref{ansatz for T(k2,p2)},\ref{ansatz for T(k2,p2) Landau Gauge}) define simple, numerically friendly {\em Ans\"{a}tze} for the transverse vertex contribution ${\cal{T}}$ to the gap equation. In Landau gauge, an extension of the {\em Ansatz} defined through Eq.~(\ref{ansatz for T(k2,p2) With Rho}) explicitly incorporates DCSB and still ensures a critical coupling independent of the $\rho$ parameter. \\

\section{Conclusions}
\label{SECTION Results and Conclusions}

In this article, we have investigated combined constraints of TTI, LKFT, MR of the massless fermion propagator, gauge-independence of the critical coupling $\alpha_c$ in quenched QED and one-loop perturbation theory in the asymptotic limit to construct a general fermion-photon vertex. We work explicitly with $Y_i$ functions, which arise naturally on the implementation of the TTI, providing, along the way, their symmetry properties
under the charge conjugation operation. Through an exact relation, we define effective $Y_i$ for which the angular dependence on the variable $q^2$ has been integrated out to make their implementation in the gap equation more efficient. As a simplifying consequence of working with $Y_i$, we observe that the kernel dependence on the Gram determinant $\nabla(k,p)$ for the mass function disappears altogether. Moreover, our study reveals that we
cannot force all $Y_i$ to be simultaneously equal to zero. It will violate the LKFT transformation law and the MR of the massless fermion propagator. We work with quite a general vertex construction~\cite{Bashir:2011dp}, formulated in terms of $Y_i$.
We also provide simple examples of this fermion-photon vertex and carry out its numerical study to compute the mass function and its variation as a function of the coupling strength. The results clearly follow Miransky scaling law and provide $\alpha_c=\pi/3$. Moreover, anomalous mass dimension $\gamma_m=1$, as has been advocated in several previous works~\cite{Holdom:1988gs,Holdom:1988gr,Bashir:1995qr,Akram:2012jq}. Also, this critical coupling is gauge independent. As
mentioned before, fermion-photon vertex enters the SDE study of several hadronic obseravbles, such as form factor calculations, where photons interact with quarks. Therefore, an improved understanding of this vertex, such as the one detailed in this article, is very important. Moreover, a natural extension of our work for the quark-gluon vertex in QCD is currently underway. \\

\noindent
{\bf Acknowledgements:} We dedicate this article to the fond memory of Prof. M.R. Pennington with whom the hunt for a gauge independent DCSB began about three decades ago. We are grateful to C. Schubert and R. Williams for helpful comments on the draft version of this article. This research was partly supported by Coordinaci\'on de la Investigaci\'on Cient\'ifica (CIC) of the University of Michoacan and
Consejo Nacional de Ciencia y Tecnolog\'ia (CONACyT), Mexico, Grant nos. 4.10 and CB2014-22117, as well as grants CNPq no. 307485/2017-0 and FAPESP no. 2016/03154-7.

%%%%%%%%%%%%%%%%%%%%%%%%%%%%%%%%%%%%%%%%%%%%%%%%%%%%%%%%%%%%%%%
%%%%%%%%%%%%%%%%%%%%%%%%%%%%%%%%%%%%%%%%%%%%%%%%%%%%%%%%%%%%%%%
%%%%%%%%%%%%%%%%%%%%     BIBLIOGRAPHY      %%%%%%%%%%%%%%%%%%%%
%%%%%%%%%%%%%%%%%%%%%%%%%%%%%%%%%%%%%%%%%%%%%%%%%%%%%%%%%%%%%%%
%%%%%%%%%%%%%%%%%%%%%%%%%%%%%%%%%%%%%%%%%%%%%%%%%%%%%%%%%%%%%%%

\bibliography{TTI-References}

\end{document}